\documentclass[11pt]{article}
\usepackage{amsmath,amsthm,amsfonts,amssymb, graphicx,mathabx,epsfig}
\usepackage{bbm}
\usepackage{authblk}
\usepackage{bm}
\usepackage{graphicx,color}
\usepackage{footmisc}
\usepackage{epstopdf}
\usepackage{dsfont}
\usepackage{pgfplots}
\usepackage{subfig}
\usetikzlibrary{fit}
\usepackage{slashed}

\font\ottorm=cmr8 scaled\magstep1 
scaled\magstep1                  
scaled\magstep1 \font\msytw=msbm10 scaled\magstep1
 
scaled\magstep1                 \font\indbf=cmbx10 scaled\magstep2

scaled \magstep2

{\count255=\time\divide\count255 by 60
\xdef\hourmin{\number\count255}
        \multiply\count255 by-60\advance\count255 by\time
   \xdef\hourmin{\hourmin:\ifnum\count255<10 0\fi\the\count255}}

\usepackage[hmargin={2.7cm,2.7cm},vmargin={2.5cm,2.5cm},centering]{geometry}
\usepackage{amssymb,amsmath,amsthm,amsfonts,amscd}
\usepackage{graphicx}
\usepackage{graphicx}% Include figure files
\usepackage{dcolumn}% Align table columns on decimal point
\usepackage{bm}% bold math
\usepackage{epstopdf}
\newcount\driver
\newcount\bozza

\font\ottorm=cmr8 scaled\magstep1 
scaled\magstep1                  
scaled\magstep1 \font\msytw=msbm10 scaled\magstep1
 
scaled\magstep1                 \font\indbf=cmbx10 scaled\magstep2

scaled \magstep2

{\count255=\time\divide\count255 by 60
\xdef\hourmin{\number\count255}
        \multiply\count255 by-60\advance\count255 by\time
   \xdef\hourmin{\hourmin:\ifnum\count255<10 0\fi\the\count255}}

\let\a=\alpha \let\b=\beta    \let\g=\gamma     \let\d=\delta     \let\e=\varepsilon
  \let\h=\eta     \let\th=\vartheta \let\k=\kappa     \let\l=\lambda
\let\m=\mu    \let\n=\nu                      \let\r=\rho
\let\s=\sigma \let\t=\tau            
\let\ps=\psi        
\let\G=\Gamma \let\D=\Delta       \let\L=\Lambda

\def\PP{{\cal P}}\def\EE{{\cal E}}\def\VV{{\cal V}}
\def\HH{{\cal H}}\def\WW{{\cal W}}
\def\TT{{\cal T}}\def\NN{{\cal N}}\def\ZZ{{\cal Z}}
\def\RR{{\cal R}}\def\LL{{\cal L}}

\def\xx{{\bf x}}
  
\def\yy{{\bf y}}\def\nn{{\bf n}}
\def\uu{{\bf u}}
 \def\bP{{\bf P}}
\def\tt{{\bf t}}\def\bT{{\bf T}}

       \def\oo{{\underline \omega}}
\def\ee{{\underline \varepsilon}}

\def\uu{\bf u}

\def\RRR{\hbox{\msytw R}}

        \def\ZZZ{\hbox{\msytw Z}}

        \def\EE{\hbox{\msytw E}}

\let\bs=\backslash
\let\==\equiv

\let\io=\infty
\let\0=\noindent

\def\*{{\hfill\break\null\hfill\break}}

\def\tilde#1{{\widetilde #1}}

\def\la{{\langle}}
\def\ra{{\rangle}}

\def\tende#1{\,\vtop{\ialign{##\crcr\rightarrowfill\crcr
             \noalign{\kern-1pt\nointerlineskip}
             \hskip3.pt${\scriptstyle #1}$\hskip3.pt\crcr}}\,}
\def\otto{\,{\kern-1.truept\leftarrow\kern-5.truept\to\kern-1.truept}\,}

\def\wh#1{\widehat{#1}}
\def\hat#1{\wh{#1}}
\def\sqt[#1]#2{\root #1\of {#2}}

\def\bp{{\bar \ps}}

\def\PP{{\cal P}}\def\EE{{\cal E}}\def\VV{{\cal V}}
\def\HH{{\cal H}}\def\WW{{\cal W}}
\def\TT{{\cal T}}\def\NN{{\cal N}}\def\ZZ{{\cal Z}}
\def\RR{{\cal R}}\def\LL{{\cal L}}

\def\T#1{{#1_{\kern-3pt\lower7pt\hbox{$\widetilde{}$}}\kern3pt}}
\def\VVV#1{{\underline #1}_{\kern-3pt
\lower7pt\hbox{$\widetilde{}$}}\kern3pt\,}
\def\W#1{#1_{\kern-3pt\lower7.5pt\hbox{$\widetilde{}$}}\kern2pt\,}

\def\indica{\leaders \hbox to 0.5cm{\hss.\hss}\hfill}
\def\guida{\leaders\hbox to 1em{\hss.\hss}\hfill}
\mathchardef\oo= "0521

\def\xx{{\bf x}}
\def\yy{{\bf y}}\def\nn{{\bf n}}
\def\uu{{\bf u}}
 \def\bP{{\bf P}}
\def\tt{{\bf t}} 
\def\oo{{\underline \omega}}

\def\qed{\raise1pt\hbox{\vrule height5pt width5pt depth0pt}}
 
  \def\bp{{\bar p}} 

\def\indic{\hbox{\raise-2pt \hbox{\indbf 1}}}

\def\RRR{\hbox{\msytw R}}

 \def\ZZZ{\hbox{\msytw Z}}

\def\ins#1#2#3{\vbox to0pt{\kern-#2 \hbox{\kern#1 #3}\vss}\nointerlineskip}

\newdimen\xshift \newdimen\xwidth \newdimen\yshift
\def\insertplot#1#2#3#4#5#6{%
\xwidth=#1pt \xshift=\hsize \advance\xshift by-\xwidth \divide\xshift by 2%
\begin{figure}[ht]
\vspace{#2pt} \hspace{\xshift}
\begin{minipage}{#1pt}
#3 \ifnum\driver=1 \griglia=#6
\ifnum\griglia=1 \openout13=griglia.ps \write13{gsave .2
setlinewidth} \write13{0 10 #1 {dup 0 moveto #2 lineto } for}
\write13{0 10 #2 {dup 0 exch moveto #1 exch lineto } for}
\write13{stroke} \write13{.5 setlinewidth} \write13{0 50 #1 {dup 0
moveto #2 lineto } for} \write13{0 50 #2 {dup 0 exch moveto #1
exch lineto } for} \write13{stroke grestore} \closeout13
\includegraphics{griglia.ps} \fi
\includegraphics{#4.ps}\fi%
\ifnum\driver=2 \fi
\end{minipage}
\caption{#5}
\end{figure}
}
\newdimen\shift \shift=-1.5truecm
\def\lb#1{%
\ifnum\bozza=1
\label{#1}\rlap{\hbox{\hskip\shift$\scriptstyle#1$}}
\else\label{#1} \fi}

\def\be{\begin{equation}}
\def\ee{\end{equation}}
\def\bea{\begin{eqnarray}}\def\eea{\end{eqnarray}}
\def\bean{\begin{eqnarray*}}\def\eean{\end{eqnarray*}}
\def\bfr{\begin{flushright}}\def\efr{\end{flushright}}
\def\bc{\begin{center}}\def\ec{\end{center}}
\def\bal{\begin{align}}\def\eal{\end{align}}
\def\ba#1{\begin{array}{#1}} \def\ea{\end{array}}
\def\bd{\begin{description}}\def\ed{\end{description}}
\def\bv{\begin{verbatim}}\def\ev{\end{verbatim}}
\def\nn{\nonumber}
\def\Halmos{\hfill\vrule height10pt width4pt depth2pt \par\hbox to \hsize{}}
\def\pref#1{(\ref{#1})}

\def\ins#1#2#3{\vbox to0pt{\kern-#2 \hbox{\kern#1 #3}\vss}\nointerlineskip}

\newdimen\xshift \newdimen\xwidth \newdimen\yshift
\newcount\griglia

\def\insertplot#1#2#3#4#5#6{%
\xwidth=#1pt \xshift=\hsize \advance\xshift by-\xwidth \divide\xshift by 2%
\begin{figure}[ht]
\vspace{#2pt} \hspace{\xshift}
\begin{minipage}{#1pt}
#3 \ifnum\driver=1 \griglia=#6
\ifnum\griglia=1 \openout13=griglia.ps \write13{gsave .2
setlinewidth} \write13{0 10 #1 {dup 0 moveto #2 lineto } for}
\write13{0 10 #2 {dup 0 exch moveto #1 exch lineto } for}
\write13{stroke} \write13{.5 setlinewidth} \write13{0 50 #1 {dup 0
moveto #2 lineto } for} \write13{0 50 #2 {dup 0 exch moveto #1
exch lineto } for} \write13{stroke grestore} \closeout13
\includegraphics{griglia.ps} \fi
\includegraphics{#4.ps}\fi%
\ifnum\driver=2 \fi
\end{minipage}
\caption{#5}
\end{figure}
}
\newdimen\shift \shift=-1.5truecm
\def\lb#1{%
\label{#1}\rlap{\hbox{\hskip\shift$\scriptstyle#1$}}
\else\label{#1} \fi}

\def\be{\begin{equation}}
\def\ee{\end{equation}}
\def\bea{\begin{eqnarray}}\def\eea{\end{eqnarray}}
\def\bean{\begin{eqnarray*}}\def\eean{\end{eqnarray*}}
\def\bfr{\begin{flushright}}\def\efr{\end{flushright}}
\def\bc{\begin{center}}\def\ec{\end{center}}
\def\bal{\begin{align}}\def\eal{\end{align}}
\def\ba#1{\begin{array}{#1}} \def\ea{\end{array}}
\def\bd{\begin{description}}\def\ed{\end{description}}
\def\bv{\begin{verbatim}}\def\ev{\end{verbatim}}
\def\nn{\nonumber}
\def\Halmos{\hfill\vrule height10pt width4pt depth2pt \par\hbox to \hsize{}}
\def\pref#1{(\ref{#1})}

\usepackage{amsmath}
\usepackage{amsfonts}
\usepackage{amssymb}
\usepackage{epstopdf}

\font\msytw=msbm9 scaled\magstep1 
scaled\magstep1

\let\a=\alpha \let\b=\beta  \let\g=\gamma  \let\d=\delta
\let\e=\varepsilon
  \let\h=\eta   \let\th=\theta \let\k=\kappa \let\l=\lambda
\let\m=\mu    \let\n=\nu             \let\r=\rho
\let\s=\sigma \let\t=\tau    
\let\ps=\Psi   
\let\G=\Gamma \let\D=\Delta  \let\L=\Lambda

\def\EE{{\cal E}} \def\VV{{\cal V}}
 \def\WW{{\cal W}}
\def\TT{{\cal T}}\def\NN{{\cal N}} 
\def\RR{{\cal R}}\def\LL{{\cal L}}

 \def\xx{{\bf x}} \def\yy{{\bf y}} 

\def\PP{{\bf P}}

\def\nn{\nonumber}

\def\RRR{\hbox{\msytw R}}

 \def\ZZZ{\hbox{\msytw Z}}

\def\\{\hfill\break}
\def\={:=}
\let\io=\infty
\let\0=\noindent

\def\tende#1{\,\vtop{\ialign{##\crcr\rightarrowfill\crcr\noalign{\kern-1pt
    \nointerlineskip} \hskip3.pt${\scriptstyle #1}$\hskip3.pt\crcr}}\,}
\def\otto{\,{\kern-1.truept\leftarrow\kern-5.truept\to\kern-1.truept}\,}

\def\wh{\widehat}
\def\to{\rightarrow}
\def\la{\left\langle}
\def\ra{\right\rangle}
\def\qed{\hfill\raise1pt\hbox{\vrule height5pt width5pt depth0pt}}

\def\be{\begin{equation}}
\def\ee{\end{equation}}
\def\bp{\begin{pmatrix}}
\def\ep{\end{pmatrix}}
\def\bea{\begin{eqnarray}}
\def\eea{\end{eqnarray}}
\def\nn{\nonumber}
\def\pref#1{(\ref{#1})}

\def\lb{\label}

\newtheorem{lemma}{Lemma}[section]
\newtheorem{theorem}{Theorem}[section]

\begin{document}

\title{Anomaly cancellation in the lattice effective electroweak theory}

\author{Vieri Mastropietro }

\affil{University of Milano, Italy}
\affil{
\texttt{vieri.mastropietro@unimi.it}}

%\texttt{vieri.mastropietro@unimi.it}

\maketitle
\begin{abstract} 
%Chiral gauge invariance, necessary
%for the Standard Model consistency, holds at a perturbative level under the anomaly cancellation conditions, 
%constraining the 
%charges of elementary particles.
The anomaly cancellation is  at the basis of the perturbative
consistence of the Standard Model and it provides a partial explanation
of charge quantization. We consider
 an effective Electroweak theory on a lattice, with a quartic interaction 
describing the weak forces and an interaction with the e.m. field. 
We prove the validity of  the anomaly cancellation 
at a non perturbative level and with a finite lattice cut-off,
even if  the lattice breaks some important symmetries, on which perturbative arguments for the cancellation are
based. The method of the proof has analogies with the one adopted for establishing
universality in transport of
quantum materials.
\end{abstract} 
\maketitle

%\section{Effective electroweak theory and main result}

\section{Introduction and Main results}

\subsection{The Electroweak theory}

The {\it Standard Model} describes the interaction of all the known elementary particles by 
three of the four fundamental forces (electromagnetic, weak and strong interactions).
The first two are described by the {\it Electroweak} sector of the theory, see e.g. \cite{O}, \cite{MS},\cite{Mai};
the e.m. forces are long ranged while the weak ones are short ranged and {\it chiral}, that is left and right handed fermions interact 
in a different way. A basic property
of the theory is its {\it perturbative} consistence or {\it renormalizability}, that is the physical observables are series which are order by order finite if 
a proper choice of the parameter is done.
Such requirement is quite restrictive and allowed one to select the number of possible theories.
Renormalizability was indeed lacking in the first theory of weak interactions,
the Fermi theory 
\cite{Fe}
(the "Tentativo"), where forces are transmitted via a current-current contact interaction. The
degree of divergence $D$ of a graph with $n_\psi$ external fermion lines
of order $n$ is $D=4-3 n_\psi/2+2 n$, where dependence on the order signal non-renormalizability.
In QED the forces are instead mediated by
a boson and the degree of divergence is, if $n_G$ are the external boson lines, 
$D=4-3 n^\psi/2-n^G$, corresponding to a renormalizable theory. 
This suggested the idea that forces are mediated by $W,Z$ 
bosons derived by a gauge principle
(Yang-Mills theory), and indeed the corresponding theory is renormalizable if the bosons are massless. Weak forces are however short ranged and bosons must be massive, and adding a mass in the action can break the renormalizability.
The massive vector boson propagator has the form
${1\over k^2+M^2}(\d_{\m\n}+ {k_\m k_\n\over M^2})$ and the second piece is non-decaying; the 
scaling dimension is 
$D=4+(2-z)n/2-3 n^\psi/2-(4-z)n^G/2$ with 
$z=0$. Even if the dimension $D$ correspond to a non-renormalizable 
theory, there can be a reduction making it renormalizable. This is what happens in QED with a massive photon breaking gauge invariance;
conservation  of current $k_\m \hat j_\m=0$, 
and the corresponding Ward Identities with dimensional regularization,
ensures  that the non decaying part of the propagator does not contribute and $z=2$ in $D$. In a chiral gauge theory this does not happen, as
the conservation of the chiral currents is broken by the fermion mass. 
A perturbatively renormalizable theory for electroweak forces 
was then proposed by Weinberg \cite{W}; the fermion and gauge boson masses are not added to the action but
generated by the interaction with a boson Higgs field which was unobserved at that time. Renormalizability was conjectured in 
\cite{W} and proved by t'Hooft 
\cite{TH}. Even if present at a classical level, the conservation of the current
is broken by the {\it anomalies} \cite{A} and the perturbative
renormalizability requires that the anomalies cancel out exactly.  Anomalies are associated to the three-current correlation, which is given
by a series of terms; it turns out that the lowest order, expressed by "triangle graphs", cancels out provided that the electric charges
verify an algebraic "anomaly cancellation condition" 
and higher order terms vanish 
due to the Adler-Bardeen property
\cite{AB}. The anomaly cancellation condition \cite{1a} 
provides an explanation of the charge quantization, that is the principle that all charges
of elementary particles are integer multiple of a quantum of charge ($1/3$ of the electron charge).  Therefore, the requirement of
renormalizability of the theory
leads to the prediction of the existence of $W,Z$ particles, of the Higgs and to explain the charge quantization.

The Standard Model gives the best understanding of the fundamental level of reality but its mathematical status is unclear. Perturbative renormalizability
only implies that the physical properties are expressed by a 
 series order by order finite, but it is unclear if such series 
are the expansion of something; they 
are most likely 
divergent and even not asymptotic, see e.g. \cite{Se}. The triviality phenomenon, rigorously proved only for $\phi^4$ models  \cite{F1},\cite{A11} 
but expected also in QED and electroweak theory, says that the theory is indeed gaussian.
The search for a non-perturbative version of the Standard Model leads to the proposal of a {\it lattice} formulation, whose step is kept
finite and acts as an ultraviolet cut-off. The physical motivation is to look at the 
Standard Model as an {\it effective theory}, see e.g. 
\cite{Th1},\cite{Po},
\cite{W1},
valid up to a certain energy scale and emerging from some more fundamental and unknown theory.
Physical consistence only requires that the cut-off
must be higher than the experiment energy scale, so that its effect is essentially invisible in particle experiments.
With this approach the counterpart of non-renormalizability or renormalizability is the size of the maximum cut-off allowed, that is the range
of validity of the theory.
With a finite cut-off,
the Fermi theory is a meaningful theory
expressed by 
{\it convergent} series,
 with radius proportional to the square
of the inverse of the cut-off; this has been proved with a momentum cut-off in \cite{M1aa}, \cite{Ma1},and with a lattice cut-off
in \cite{GMaa} and \cite{M2aa} (in the context of condensed matter models which are essentially equivalent to QFT on a lattice), using 
the fermionic Renormalization Group, see e.g. \cite{Mbook}.
% and in particular Battlle-Brydges-Federbush formulas \cite{B}, tree expansions %\cite{Ga} and determinat bound
%\cite{GK},\cite{Le}.
The non-perturbative validity of lattice Fermi theory is valid up to energies
(much) smaller than $m_W\sim 89 Gev$,
an energy range typical of experiments before the Eighties.
 The perturbative renormalizability
of the Standard Model suggests that a non-perturbative construction of the electroweak sector with a finite cut-off is possible up
to scale exponentially high in the inverse coupling, a value $\sim e^{286} Ev$ much higher even than the Plank scale. This
exponentially high result is suggested both from perturbative lowest order computations (the solution of the Renormalization Group flow
equations describing the effective coupling is $O(1)$ only close to such scale, see e.g.  \cite{MS}) and in general from 
perturbative dimensional arguments. Nevertheless, even the lattice formulation of the electroweak theory (leaving outside the mathematical control of functional integrals)
has proven to be extremely hard, see e.g. \cite{B3c}- \cite{D}. One needs that the anomaly cancels with a finite lattice under the same condition
on the charges as in the continuum, in order to have an improvement in the degree of divergence as in the perturbative case.
The argument holding in the continuum relies on relativistic and chiral symmetries which are
broken by the lattice
\cite{NN}; in particular if one requires to get the correct continuum limit a Wilson term has to be added, coupling the
$L$ and $R$ part of fermions
breaking chiral symmetry.
Even if the terms due to the lattice are irrelevant in the RG terms,
formally vanishing in the limit, at finite cut-off they can nevertheless add finite contributions breaking the cancellation.
More elaborate lattice constructions
\cite{B3}, \cite{B4} claim anomaly cancellation but 
are based on an order by order analysis 
and non-perturbative contributions cannot be excluded.
Functional integral arguments, extending the approach in \cite{F}, are tipically one loop results. Topological arguments
work with classical gauge fields
\cite{Z}, but 
but in presence of the interaction they require massive fermion, as for topological insulators \cite{HM}. 
While there is no non-perturbative proof of the anomaly cancellation on a lattice,
support for it comes from 
apparently unrelated properties of quantum materials.
Dirac semimetals can be considered as condensed matter realization 
of interacting Dirac fermions on a lattice \cite{K}
and transport coefficients are related to anomalies \cite{SS},\cite{SS1}.
Recent experiments in Graphene \cite{Na} showed that
the optical conductivity, related to the parity anomaly of the emerging $d=2+1$ QFT, is universal and
equal to its non interacting continuum value; 
lattice or interactions, which are surely present in such materials, do 
not modify the anomaly. This experimental result has been rigorously explained 
in the case of weak Hubbard interactions in 
\cite{BGM}. Universality of the Chiral anomaly in $3+1$ dimensional Weyl semimetals 
has been also established \cite{BGM1}. It is therefore natural to apply such methods to latttice QFT.
 
In the electroweak theory 
one has a family of particles composed by two fermions (the electron $\e$ and the neutrino
$\n$) and two quarks, the up $u$ and down $d$). Two regimes can be identified.
A first high energy regime from the cut-off up to $W,Z$ mass scale generated by the Higgs, expressed in terms of the gauge fields $W^k_\m$, $k=1,2,3$ and
$B_\m$ (associated to $SU(2)$ and $U(1)$ invariance)
and the fermionic fields $\psi_i$.  The current associated to $B_\m$ is
$j_\m^B=\sum_{i=e,\n,u,d} \sum_{s=L,R}  Y_{i,s}/2 \bar\psi_i\g_\m(1+\e_e \g_5)\psi_i$ where $\e_L=-\e_R=1$, 
$Y_{i,s}$ is the hypercharge of the $i$-particle with chirality $s$
and 
; the current associated to $W^k_\m$ is
$j^{k,W}_\m= \bar\Psi_{l,x}\t^k \g_\m \Psi_{l,x}+\bar\Psi_{l,x}\t^k \g_\m \Psi_{l,x}
  $ with $\t^k$ pauli matrices,
$\Psi_{l,x}=(\psi_{\n,L,x},\psi_{\e,L,x} )$ (the leptonic doublet) and $\Psi_{q,x}=(\psi_{u,L,x},\psi_{d,,x} )$ (the quark doublet).
The $W^\pm_\m$ becomes massive due to the interaction with the Higgs and the $W^3_\m, B_\m$ are combined in a massive $Z_\m$ field and massless $A_\m$. The {\it anomaly} is the non-vanishing value of
$p_\m <j_\m^{B,A}j_\n^{B,V}j_\r^{B,V}>$, where $j_\m^{B,A}$ and $j_\n^{B,V}$ are the axial and vector part of the current, and it signals the 
non-conservation of the current (in the classical case it would be vanishing by Noether theorem).
In the second lower energy regime from the $W,Z$ mass scale to zero, 
the non-local nature of the $W,Z$ interaction cannot be detected and an effective description is valid, of the form
\be
\sum_\m [{g^2\over 2 M_W^2}  j_{\m,x}^{W+} j^{W-}_{\m,x}+{\bar g^2\over 2  M_Z^2}
j_{\m,x}^Z j^Z_{\m,x}+e A_{\m,x} j_{\m,x}^{e.m.}]\label{e.m.1}
\ee 
We introduce a lattice version of the effective electroweak model \pref{e.m.1}
inspired by \cite{B3c}, \cite{B3d}, with a Wilson term, square lattice
and boson mass. The volume is sent to infinity.
Our main result can be informally stated as follows.
\vskip.3cm
{\it The correlations of the lattice electroweak theory are given by convergent series ("expansion of anything") for an inverse lattice step of the order of boson mass and the anomaly is vanishing up to subleading corrections under the condition
$\sum_i (Y_{i, R}^3-Y_{i, L}^3)=0$.}
\vskip.3cm

The vanishing of the anomaly, obtained up to now only at a purely perturbative level, is proved 
with a finite lattice cut-off, even if the cut-off breaks important symmetries on which the perturbative cancellation 
were based, like the Lorentz or the chiral one. The anomaly vanishes if
\be
(Y_{e,L})^3+(Y_{\n,L})^3+3 (Y_{u, L})^3+3(Y^{d, L})^3
-(Y_{e, R})^3-(Y_{\n, R})^3-3 (Y_{u, R})^3-3(Y^{d, R})^3=0\label{assa}
\ee
which is indeed verified by particles in nature; 
electric charges are given by $Q_i={\e\over 2}(\t_{i, s}+Y_{i, s})$, 
$\t_{e, L}=\t_{d, L}=-1$, $\t_{\n, L}=\t_{e, L}=1$, $\t_{i, R}=0$,
and
$Y_{\n, L}=Y_{e, L}=-1$, $Y_{u, L}=Y_{d, L}=1/3$,
$Y_{\n, R}=0$, $Y_{e, R}=-2$, $Y_{u,R}=4/3$, $Y_{d, R}=-2/3$
so that $6(1/3)^3+2(-1)^3-3(4/3)^3-3(-2/3)^3-(-2)^3=0$. 
The lattice cut-off plays an essential role as it ensure the validity of Ward Identities; by other regularizations, like the momentum one, only a partial cancellation can be proved \cite{Ma1}.
The method of the proof has analogies with the one adopted for establishing
universality in transport of
quantum materials \cite{BGM},\cite{BGM1}. The vanishing of the anomaly with a finite lattice is established for a lattice cut-off of the order of boson mass;
its validity for such cut-off  
is a prerequisite for its validity at higher values, where is an essential
ingredient for the (possible) construction of the theory up to exponentially high scales. 

In trying to go to an higher cut-off one needs to perform also a decomposition in the boson fields (replacing the effective interaction with the gauge one), while in the present analysis the multiscale analysis is only for fermions. The anomaly cancellation and the associated Ward Identities 
should ensure the decrease of the degree of divergence (from non-renormalizable to renormalizable, with massive boons and massless fermions), even in absence of full gauge invariance; a (non chiral) realization of this property in $d=1+1$ at a non perturbative level is in the Sommerfield model
\cite{M22} (from renormalizable to superrenormalizable).
Moreover
one needs to use Ward Identities at each Renormalization Group iteration as done for the renormalizable $d=1+1$ Thirring model \cite{BFM1}, to control the flow of the effective couplings.

\subsection{The model}

%We define the doublets
%$\Psi^\pm_{l,x}=(\psi^\pm_{\n,L,x},\psi^\pm_{e,L,x})$ and  $\Psi^\pm_{q,x}=(\psi^\pm_{u,L,x},\psi^\pm_{d,L,x})$. 

%The effective Electroweak model describes a family of fermions, composed by %two massless leptons $(\n,e)$ and two quarks $(u,d)$. 
%The electromagnetic interaction is through a field $A_\m$ while the weak %forces are described by a Fermi current-current interaction,
%which is the low energy limit of the gauge interaction with $W,Z$ bosons. The %cut-off is introduced via a lattice
%and the photon field via the Peierls substitution. 

We consider 
a set of {\it Grassmann variables} $\psi^\pm_{i,s,j,x}$ with 
\begin{enumerate}
\item $x\in \L$ is the coordinate,  $\L=[0,L]^4\cap a \ZZZ^4$, $L=N a$ with $N$ integer;
\item $i \in (\n_1,e_1,u_{1,c},d_{1,c},\n_2,e_2,u_{2,c},d_{2,c})
$ is the {\it particle index} and $i_1\in (\n_1,e_1,u_{1,c},d_{1,c})$, or $i_2\in (\n_2,e_2,u_{2,c},d_{2,c})$; 
\item $c$ is the {\it colour index}
 $c\in (r,g,b)$ ;
\item
 $s\in (L,R)$ is the {\it chiral index};
\item $j\in (1,2)$ the component index
\end{enumerate}
Anti-periodic boundary conditions are imposed.
We use also the notation $\psi^\pm_{i,x}=(\psi^\pm_{i,L,x},\psi^\pm_{i,R,x})$ and 
$\psi^\pm_{i,s,x}=(\psi^\pm_{i,s,1,x}, \psi^\pm_{i,s,2,x})$. 
The {\it fermionic integration} is
\be
P(d\psi)={1\over\NN_\psi}  [\prod_{i,x} d\psi^+_{i,x}  d\psi^-_{i,x}] 
e^{-S}
\ee where $\NN_\psi$ is a normalization constant and
%$\int dx\equiv a^4\sum_x$
%
\bea
&&S={1\over 2 a}  a^4\sum_x \sum_i[ \sum_{\m=0}^3 (  
 \psi^+_{i,x}\g_0\g_\m   \psi^-_{i, x+a_\m}-\psi^+_{i, x+a_\m}\g_0\g_\m   \psi^-_{i, x})+\nn\\
&&r( \psi^+_{i,x}\g_0 \psi^-_{i, x+a_\m}+\psi^+_{i,x+a_\m }\g_0 \psi^-_{i, x}
-\psi^+_{i, x}\g_0\psi^-_{i,x}) ]\label{jkkh}
\eea
with $a_0=(a,0,0,0)$,...,$a_3=(0,0,0,a)$, the gamma matrices are $\g_\m,\g_5$, $\m=0,1,2,3$ 
$\g_0= \begin{pmatrix} 0 & I \\ I &0 \end{pmatrix}\quad \g_j= \begin{pmatrix} 0 & i\s_j \\-i\s_j &0 \end{pmatrix}, \quad\g_5=\begin{pmatrix}&I&0\\
          &0&-I\end{pmatrix}$.
We set $\s_\m^L=(\s_0,i \s) $ e $\s_\m^R=(\s_0,-i \s) $ and
the matrices $\s_1=\begin{pmatrix}&0&1\\
          &1&0\end{pmatrix}
 \quad
\s_2=\begin{pmatrix}
&0&-i\\ &i&0
\end{pmatrix}
\quad\s_3=\begin{pmatrix}&1&0\\
          &0&-1\end{pmatrix}$. 
We equivalently write $S$
as
\bea
&&S={1\over 2 a} \sum_{i}
a^4 \sum_x [
\sum_{s}\sum_{\m}
 (\psi^+_{i,s, x}\s_\m^s   \psi^-_{i, s, x+a_\m}-\psi^+_{i,s, x+a_\m}\s^s_\m \psi^-_{i, s, x})+ r( \psi^+_{i, L, x}\psi^-_{i, R, x+a_\m }+\\
&&\psi^+_{i, L, x+a_\m } \psi^-_{i,c,R, x}
-2\psi^+_{i, L, x}\psi^-_{i,c, R, x} 
 +\psi^+_{i,R, x}\psi^-_{i, L, x+a_\m}+\psi^+_{i,R, x+a_\m} \psi^-_{i,L, x}
-2\psi^+_{i,R, x}\psi^-_{i,L, x}) ]\nn\label{s}
\eea
The Fermi interaction is given by 
\be
V_F= {g^2\over 2 M_W^2} \sum_\m
 a^4 \sum_x  j_{\m,x}^{+W}  j_{\m,x}^{-W} +{\bar g^2\over 2 M_Z^2} \sum_\m
a^4 \sum_x j_{\m,x}^Z  j_{\m, x}^Z\label{quar} 
\ee
where the charged weak currents are
\be
j_{\m,x}^{+W}=
\psi^+_{\n_1, L,x}\s_\m^L \psi^-_{e_1,L,x}+ 
\sum_c \psi^+_{u_{1,c},L,x}\s_\m^L\psi^-_{d_{1,c},L,x};\quad  
j_{\m,x}^{-W}=
\psi^+_{e_1, L,x}\s_\m^L\psi^-_{\n_1,L,x}+\sum_c 
\psi^+_{u_{1,c},L,x}\s_\m^L\psi^-_{d_{1,c},L,x}
\ee
 and the neutral current is
\be
j^Z_{\m,x}=\sum_{i_1} (1-\sin^2 \th Q_i)
\psi^+_{x,i_1,L}\s_\m^s \psi^-_{x,i_1,L}+\sum_{i_2} (-1-\sin^2 \th Q_i)
\psi^+_{x,i_2,R}\s_\m^s \psi^-_{x,i_2,R}
\ee
with $\cos \th=g/g'$, $\bar g=\sqrt{g^2+g'^2}$.
The fermionic mass counterterm is
\be
V_c=  \sum_{i} a^{-1}  \n_i  a^4 \sum_x (\psi^+_{i,L,x}\psi^-_{i,R,x}+\psi^+_{i,R,x}\psi^-_{i,L,x})\ee
We introduce the e.m. field
$A_\m(x): \L\to \RRR^4$ with periodic boundary conditions;
the bosonic integration is
\be
P(dA)={1\over \NN_A} [\prod_{x\in\L} \prod_{\m=0}^3 d A_\m(x)] e^{-{ 1\over 2} a^4\sum_x \sum_{\m=0}^3
A_\m(x) (-\D+M^2)A_\m(x) }
\ee
where 
$\NN_A$ is the normalization, $\D f= {1\over a^2}\sum_{\m=0}^3 (f(x+a_\m)+f(x-a_\m) -2 f(x)) $.

The 
{\it generating function} is
\be
e^{\WW_\L(J,J^5,\phi)} =\int P(dA)\int P(d\psi) e^{V_c(\psi)+ 
V_{e.m.}(\psi,A,J)+V_F(\psi) +B(\psi,J^5,\phi)}\label{llll}
\ee
where
\bea
&&V_{e.m.}(\psi,A,J)={1\over 2 a} \sum_{i}
a^4 \sum_x [
\sum_{s}\sum_{\m}
 (\psi^+_{i,s, x} G^+_{\m,i,s}\s_\m^s   \psi^-_{i, s, x+a_\m}-
\psi^+_{i,s, x+a_\m} G^-_{\m,i,s}\s^s_\m \psi^-_{i,s, x})+\\
&& r(
 \psi^+_{i, L, x} \tilde G^+_{\m,i}\psi^-_{i, R, x+e_\m a}+\psi^+_{i, L, x+a_\m }\tilde G^-_{\m,i} \psi^-_{i,R, x} 
 +\psi^+_{i,R, x}\tilde G^+_{\m,i}
\psi^-_{i, L, x+a_\m}+\psi^+_{i,R, x+a_\m} \tilde G^-_{\m,i}\psi^-_{i,L, x}
]\nn\label{sd}
\eea
with
\be
G^\pm_{\m,i,s}=
a^{-1}( :e^{\pm i a ( \e Q_i  b_{i,s}   A_\m(x) +
Y_i J_{\m,x}  )}:-1)\quad\quad  \tilde G^\pm_{\m,i}=
a^{-1}( e^{\pm i a 
Y_i J_{\m,x}  }-1)
\ee
and 
\be
 b_{i_1,L}=b_{i_2,R}=1\quad\quad  b_{i_1,R}=b_{i_2,L}=0
\ee
with $Q_{i_1}=Q_{i_2}$ and 
$:e^{\pm i \e Q_i  b_{i, s} a A_\m(x)
 }:=e^{\pm i \e Q_i  a b_{i, s}  A_\m(x)
 } e^{{1\over 2}
(\e Q_i)^2  a^2  b_{i, s}^2  g^A_{\m,\m} (0,0)) }$.
The source term is
\be
B(\psi,J^5,\phi)=a^4\sum_x 
[
j^5_{\m,x} J^5_{\m,x}\nn\\
%+\sum_{j=l,q}\sum_{k=1}^3 J^W_{\m,j,k, x} \bar j^W_{\m,j,k x}
+\sum_{i,s}  (\psi^+_{i,s,x} \phi^-_{i,s,x}+\psi^-_{i,s,x} \phi^+_{i,s,x})]
\label{llll1}
\ee
where
\be
j^5_{\m,x}=\sum_{i,s} \tilde \e_i \e_s Y_i Z_{i,s}^5 \psi^+_{x,i,s}\s_\m^L \psi^+_{x,i,s}
\ee
with $\tilde \e_{i_1}=-\tilde \e_{i_2}=1$ and $\e_{L}=-\e_{R}=1$. 
We assume that the hypercharges $Y_i$ are such that
\be
Y_{\n_1}-Y_{e_1}=Y_{u_1}-Y_{d_1}\label{cond1}
\ee
%
%so that the
%generating function is invariant under the  transformation
%$
%\psi^+_{i,c, s,x}\to \psi^\pm_{i,c, s,x} e^{\pm i e Q_i \a} $
%
%with $J, J^5$ unchanged; we also assume that 
%$
%Q_i\not= Q_j$
%for any $i,j$. 

The {\it Schwinger functions} are derivatives of the generating function. The fermionic 2-point function is
\be
S^\L_{i,s,s'}(x,y)
={\partial^2 \over \partial \phi^+_{i,s,x}\partial \phi^-_{i,s',y} }
\WW_\L(J,J^5,\phi)
|_0\label{h1}
\ee
where $|_0$ means that all the external fields are set to zero, 
and the Fourier transform is $\hat S^\L_{i,s,s'}(k)=a^4 \sum_x  S^\L_{i,s,s'}(x,0) e^{-i k x}$.
The vertex functions are 
\be
\G^\L_{\m,i',s}(z,x,y) =
{\partial^3 \WW_\L(J,J^5,\phi)\over \partial J_{\m, z}
\partial\phi^+_{i',s,x}\partial \phi^-_{i',s,y} }|_0\quad \G^{5,\L}_{\m,i, s}(z,x,y) =
{\partial^3 \WW_\L(J,J^5,\phi)\over \partial J^5_{\m, z}
\partial\phi^+_{i',s,x}\partial \phi^-_{i',s,y} }
|_0\ee 
%and $\G^{W,\L}_{\m, j, i, i',s}(z,x,y) =
%{\partial^3 \over \partial J^W_{\m,j,z}
%\partial\phi^+_{i,s,x}\partial \phi^-_{i',s,y} }
%\WW_\L(J,J^5,J^W,\phi)|_0, j=l,q$ (weak vertex).
The Fourier transform is 
\be
\hat \G^\L_{\m,i',s}(k,p) = a^4 \sum_x a^4 \sum_y
S_{2,i,s}(x,0) e^{-i p z-i k y}  \G^\L_{\m,i',s}(z,0,y)
\ee
and similarly is defined $\hat \G^{5,\L}_{\m,i,i' s}(k,p) $. The three-current 
vector and chiral correlations are
\be
\Pi^{\L}_{\m,\n, \r}(z,y,x)={\partial^3 \WW_\L\over \partial J_{\m,z}
\partial J_{\n,y}\partial J_{\r,x}}|_0\quad\quad
\Pi^{5,\L}_{\m,\n, \r}(z,y,x)={\partial^3 \WW_\L\over \partial J^5_{\m,z}
\partial J_{\n,y}\partial J_{\r,x}}|_0
\ee
%
%with $\e_{i_1}=-\e_{i_2}=1$. 
We define $\lim_{L\to\io} S^\L_{i,s,s'}(x,y)=S_{i,s,s'}(x,y)$
and similarly for the other correlations.

%Finally the wave function renormalization is defined as (the definition ils well posed by symmetry)
%
%\be
%Z^D_{i,s} I=\lim_{k\to 0} {\hat g_{i,s,s} (k)\over \hat S_{i,s,s}(k) }\label{isss} \ee
%
%where $I=\begin{pmatrix}&1&0\\
 %         &0&1\end{pmatrix}$; note that the definition is 
%and the renormalized electric charge as
%
%\be
%e^D_{i,s} \s^s_\m=
%\lim_{k,p\to 0}  { 1\over Z^D_{i,s}}\hat S_{i,s,s}(k)^{-1} \hat \G_{\m,i;i,s}(k,p)
% \hat S_{i,s,s}(k+p)^{-1}
%\label{ell}
%\ee
The parameters $\n_i, Z^5$ must be chosen to ensure the validity of suitable renormalization conditions.
The amputated vertex functions are defined as
\bea
&&\g_{\m,i,s}(k,p)=\hat S^{-1}_{i,s,s}(k)\hat \G_{\m,i,s}(k,p)\hat S^{-1}_{i,s,s}(k+p)\nn\\
&&
\g^5_{\m,i,s}(k,p)=\hat S^{-1}_{i,s,s}(k)\hat \G^5_{\m,i,s}(k,p)\hat S^{-1}_{i,s,s}(k+p)
\eea
The parameters $Z^5_{i,s,
s'}$ must be chosen so that they are the same, that is
\be
\lim_{k,p\to 0}  {\g_{\m,i,s}(k,p)\over \g^5_{\m,i,s}(k,p)}=\tilde\e_i \e_s I_2\label{rac}
\ee
with $\e_L=-\e_R=1$ and $I_2=\begin{pmatrix}
&1&0\\ &0&1
\end{pmatrix}
$. The above condition says that the charges
associated to the vector and axial current are the same, a condition that has to be imposed also in a perturbative context \cite{A}.
Finally the mass counterterms $\n_i$ must be chosen so that  the theory is massless, that is
\be
\lim_{k\to 0} S_{i,s,s}(k)|_{i,j}=\io\label{45} 
\ee
\subsection{Formal continuum theory}
 
The term proportional to $r$ in \pref{jkkh}
is the Wilson term. Its role is crucial in eliminating 
unphysical poles in the lattice propagator by introducing a coupling
between the $L$ and $R$ fermions. 
%Let us  first the interaction with the e.m. and neutral currents, that i $g=0$.
In the formal continuum limit $a \to 0$ the Wilson term disappears and 
$S+V_{e.m.}$ becomes equal to, if $J=0$
\bea
&&\int dx 
\sum_{i_1} [
\psi^+_{i_1, L,x} \s_\m^L (\partial_\m+\e  Q_{i_1}  A_\m)
\psi^-_{i_1, L,x}+\psi^+_{i_2, R,x} \s_\m^R  \partial_\m
\psi^-_{i_1, R,x}]+\nn\\
&&\int dx \sum_{i_2} [
\psi^+_{i_2, L,x} \s_\m^L \partial_\m
\psi^-_{i_2, L,x}+\psi^+_{i_2, R,x} \s_\m^R ( \partial_\m+\e  Q_{i_2}  A_\m)
\psi^-_{i_2, R,x}]
\eea
The $R$ fermions of kind $i_1$ and 
the $L$ fermions of kind $i_2$ decouple in the limit and can be ignored;
they are non interacting fictitious degrees of freedom useful in the lattice regularization, see \cite{B3c}, \cite{B3d}. Therefore in the continuum limit
$i_1$ are the left handed components and $i_2$ the right handed of the leptons and quarks, $Q_{i_1}=Q_{i_2}$.
The 
$B$ current is $
j^B_{\m,x}=\sum_{i_1}  Y_i \psi^+_{x,i_1,L}\s_\m^L \psi^+_{x,i_1,L}+\sum_{i_2}  Y_{i_2} \psi^+_{x,i_2,R}\s_\m^R \psi^+_{x,i_1,R}
$
and
vector and axial currents are
\be
2j_{\m,x}^{B,V}=\sum_{i_1} Y_{i_1} j_{\m,i_1,x}+\sum_{i_2} Y_{i_2} j_{\m,i_2,x}\quad\quad 2j_{\m,x}^{B,A}=\sum_{i_1} Y_{i_1} j^5_{\m,i_1,x}-\sum_{i_2} Y_{i_2} j^5_{\m,i_2,x}
\ee
as $2 \psi^+_{i,s}\s_\m^{s}
\psi^-_{i,s}=\bar\psi_i \g_\m (1+\e_s \g_5)\psi_i$, $\e_L=-\e_R=1$.
In the formal continuum limit $8\Pi^{\L}_{\m,\n, \r,}(z,y,x)$
corresponds to  $\la j_{\m,x}^{B,V}j_{\n,x}^{B,V} j_{\r,x}^{B,V}\ra$
and $8\Pi^{\L.5}_{\m,\n, \r,}(z,y,x)$
to $\la j_{\m,x}^{B,A}j_{\n,x}^{B,V} j_{\r,x}^{B,V}\ra$.

%The quantum analogue of such conservation 
%is the vanishing of the current correlations $\partial_\m < j_\m^B  j_{\m_1}%^B... j_{\m_n}^B   >$, which can be seen as the response of the current
%to the presence of a classical field coupled to the current itself.
%The anomaly the 3-point correlation
%$\partial_\m < j_\m^B  j_{\m_1}^B j_{\m_2}^B   >$ is however non %vanishing
%for generic values of the hypercharges; more exactly is generically non %vanishing
%the axial three current correlation $\partial_\m < j_\m^{B,A}  
%j_{\m_1}^{B,V} j_{\m_2}^{B,V}>$ unless the anomaly cancellation condition &holds. The vanishing follows
%by explicit lowest order computation (triangle graph)
%and Adler-Bardeen \cite{AB} non renormalization property.

\subsection{Ward Identities}

By performing the change of variables
\be
\psi^\pm_{i,x}\to
\psi^\pm_{i,x} e^{\pm i Y_i \a_{x} }\quad\quad
Y_{\n_1}-Y_{e_1}= Y_{u_1}-Y_{d_1}
\label{wis1}
\ee
we get 
\be
W(J,J^5,\phi)=
W(J+d \a ,J^5, e^{i Y \a}
 \phi)\label{wi}
\ee
where $J+d\a$ is a shorthand for $J_{\m,x}+d_\m \a_{x}$,
$d_\m f(x)=(f(x+a_\m)-f(x))/a$ 
and $e^{i Y\a}\phi$ is a shorthand for $e^{\pm i Y_i\a_{x}} \phi_{i, x}^{\pm}$.
Note that the quartic interaction is left invariant by this transformation.
By differentiating 
with respect to $\a_z,\phi^+_y,
\phi^-_z$ and passing to Fourier transform
we get the Ward Identities, if $\s_\m(p)=(1-e^{i p_\m a})/a$ 
and 
\be
\sum_\m \s_\m(p) \hat\G^\L_{\m,i,s}(k,p) 
= Y_i
(\hat S_{i,s,s}(k)-\hat S_{i,s,s}(k+p))\label{wia}
\ee
and 
\be
\sum_\n \s_\n(p_1+p_2)
\hat \Pi_{\m,\n,\r}(p_1,p_2)=0\label{xzx}
\ee
expressing the conservation of the current; similarly
\be
\sum_\n \s_\n(p_1)
\hat \Pi^5_{\m,\n,\r}(p_1,p_2)=\sum_\r \s_\r(p_2)
\hat \Pi^5_{\m,\n, \r}(p_1,p_2)=0\label{wib}
\ee
\subsection{Main result}

In the following we prove the following result. We set $\bar M=\min(M,M_W,M_Z)$.

\begin{theorem}   
There exists $\e_0$ such that,  
for $\e^2,g^2, \bar g^2\le \e_0 ({\bar M a})^2$, $\bar M a>1$
it is possible  to choose $\n_i,Z^5_{i,s}$ such that 
the limit $L\to\io$ of the Schwinger functions exists,
\pref{rac} and \pref{45} hold and, if $p=p_1+p_2$, $\th\ge 1/2$
\be
\sum_{\m_1}\s_{\m_1}(p)\hat \Pi^5_{\m_1,\m_2,\m_3}(p_1,p_2)=
{1\over 2\pi^2}\e_{\m_2,\m_3,\a,\b}p^1_\a p^2_\b
 [\sum_{i_1} Y_{i_1}^3- \sum_{i_2} Y_{i_2}^3]
+O(a^\th p^{2+\th})
\label{33aa}
\ee
\end{theorem}
\vskip.3cm
The lattice effective model is well defined, in the sense that correlations
are expressed by convergent expansions uniformly in the volume and
up to a inverse lattice step of the order of the mass of the gauge fields. 
The choice of $\n$ ensures that the theory is massless, that is fermionic Schwinger have a power law decay.
The presence of the lattice breaks the Lorentz and chiral symmetry of the theory.
Lorentz symmetry is recovered for energies far from the cut-off; in particular,
see below, the fermionic 2-point function $S_2(k)$ is equal to the relativistic one
up to small corrections for momenta far from the cut-off and a finite wave function renormalization, depending on the particle and chirality. 
While 
the current is conserved \pref{xzx}, by \pref{33aa} we see that the axial
current is generically non conserved, unless $[\sum_{i_1} Y_{i_1}^3- \sum_{i_2} Y_{i_2}^3]=0$; noting that in the continuum limit the particles with label $i_1$
are the left ones and with $i_2$ the right ones, one recovers the condition \pref{assa}.
The anomaly therefore vanishes under the same condition found in the continuum, with a finite inverse step of the order the order of the boson masses,
even if the lattice breaks a number of symmetries on which the cancellation in the continuum limit was based. The $O(a^\th p^{2+\th})$
are a bound on the subleading corrections, which does not exclude 
that also such terms are indeed vanishing.

The rest of this paper is organized in the following way. In \S 2 we integrate the $A$ fields reducing to a fermionic theory.
In \S 3 the fermionic sector is 
analyzed by multiscale Renormalization Group using a tree expansion
and determinant bounds for fermions.
In \S 4 the flow of running coupling constants is analyzed and finally
in \S 5 the cancellation is established; in the appendix some properties of truncated expectations are recalled.

\section{Integration of the e.m. field}

The {\it bosonic simple expectation} is defined as
\be
\EE_A(A_{\m_1}(x_1)...  A_{\m_n}(x_n))=\int P(dA) A_{\m_1}(x_1)...  A_{\m_n}(x_n)
\ee
and is expressed by the Wick rule with covariance
\be
g^A_{\m,\n}(x,y)=\d_{\m,\n} 
{1\over L^4}\sum_k   { e^{i k (x-y)} \over c(k)^2+M^2}
\ee
with $c^2(k)= \sum_{\m} (1- \cos k_\m a  ) a^{-2}$, $k=2\pi n/L$ and $k\in [-\pi/a,\pi/a]^4$, $n_i=-L/a, (L-a)/a$. 
We will need also formulas for the exponentials which are given by
\be
\EE_A(\prod_ {j=1}^n e^ {i b_j A_{\m_j}(x_j)})=e^ {-{1\over 2}\sum_ {j,j'}  
b_{j}  b_{j'} g^A_{\m_i,\m_j}(x_j,x_{j'}) }\label{sim}
\ee
We integrate the $A_\m$ fields obtaining
\be
e^{ V_A (\psi,J)} =\int P(dA) e^{V_{e.m.}(\psi, A,J) }
%=e^{\sum_{n=0}^\io{1\over n!} E^T(V_{e.m.};n) }
\label{llllw}
\ee
Note that using the notation $\int dx=a^4\sum_x$, $\a=\pm$
\bea
&&V_A (\psi,J)=\sum_{n=1}^\io {1\over n!} \EE^T_A(V_{e.m.}; ...; V_{e.m.})=\int dx \sum_{i,\a} a^{ -1}  (e^{i a \a Y_i J_{\m}(x)}-1)
O^\a_{\m,i}(\psi)+\nn\\
&&\sum_{n=2}^\io  \sum_{\underline \a, \underline\m,\underline i}   {1\over n!}\int dx_1...\int dx_n a^{-n}
 [\prod_{j=1}^n O^{\a_j}_{\m_j,i_j}
(\psi)e^{i\a_j a Y_{i_j} J_{\m_j}(x_j)}\\
&&\EE^T_A( : e^{i\a_1 b_{i_1,s_1}   \e Q_1  a A_{\m_1}(x_1) }:  ;...; :e^{i \a_n b_{i_n,s_n} \e Q_n a   (A_{\m_n
}(x_n ) }:)\nn
\eea
where $\EE^T_A$ is the truncated expectation, defined as 
$\EE^T_A(O;..;O)={\partial^n\over \partial \l^n} \log \int P(dA) e^{\l O}|_0$
(for $O$ such
that the integral is well-defined) and $O^{+}_{\m}(\psi)=\psi^+_{i,s, x}\s_\m^s   \psi^-_{i, s, x+a_\m}$ and $O^{-}_{\m}(\psi)=-
\psi^+_{i,s, x+a_\m}\s^s_\m\psi^-_{i,s, x}$. Note that the only non-vanishing terms are such that all the $b_{i,s}$ in the truncated expectations are $=1$.
The above expression can be rewritten as, properly defining the kernels
$W_{n,m} $
\be
V_A (\psi,J)=
\sum_{l,m\ge 0\atop l+m\ge 1}  \sum_{\underline \a, \underline\m,\underline i}   \int d\underline x    
[\prod_{j=1}^n O^{\a_j}_{\m_j,i_j}(\psi)] [\prod_{j=1}^m  J_{\m_i }(x_i)) ]
W_{n,m} (\underline x)\label{sas3}
\ee
%

%
%\be
%\bar
%V_{e.m.}
% (A+J)
%=\sum_{i}
%a^{-1}\int dx \sum_{j=1}^3
%\psi^+_{i_j,s_j, x_j}\s_\m^{\e_j}  G^{s_j}_{\m,i}(A+J_i)  \psi^-_{i'_j ,s'_j, x'_j}
%\ee
%
\begin{lemma} The kernels in \pref{sas3} verify the following bound, for $n\ge 2$
\be
{1\over L^4}\int d\underline x 
|W_{n,m} (\underline x)|
\le C^na^{-(4-3 n-m)} \e^{n} (M a)^{2-2n} 
\ee
%where $\d$ is the length of the shortest tree conecting all points $\underline x,\underline y$.
\end{lemma}
%Note that the bound is the dimensional one; nd extr $J$  crries  $a$  more so that $D=4-3 n_\psi/2-m$
\vskip.3cm
\noindent
{\bf Proof.} 
By definition $:e^{\pm i \e Q_i  a b_{i,s} A_\m(x )
 }:=e^{\pm i e Q_i  a A_\m(x)
 } e^{{1\over 2}
(\e Q_i)^2  a^2  b^2_{i,s} g^A_{\m,\m} (0,0)) }$ and $e^{{1\over 2}
(\e Q_i)^2  a^2  b^2_{i,s} g^A_{\m,\m} (0,0)) }$ is bounded by a constant as $|g^A_{\m,\m} (0,0)|\le C a^{-2}$.
By \pref{sim} we can write, $\a_j=\pm$
%, and the simple expextation is
%
%\be
%E_A(\prod_i e^{i \e_i e Q_i  \int_0^a ds  (A_{\m_i}
%+J_{\m_i}) (x+t e_{\m_i} ) })
%)=\d_{\sum_i \e_i,0} e^ {-{1\over 2}\sum_ {i,j}  \e_i \e_j \a_i \a_j g^A_{\m_i,\m_j}(x_i,x_j) }\label{sim}
%\ee
%
\be
\EE_A(\prod_{j\in X} e^{ i \e \a_j  Q_j   a  A_{\m_j}(x_j ) })
=e^{-V(X)}\quad V(X)={1\over 2}\sum_{j,j'\in X} V_{j,j'}\label{dl1}
\ee
where $X=(1,..,n)$ and, if $\b_j=\a_j  Q_j $
\be
V_{j,j'}= \e^2 \b_{j} \b_{j'} \EE(a A_{\m_j} (x_j) a
A_{\m_{j'}}(x_{j'}) )=
\e^2 \b_{j} \b_{j'}
a^2  g^A_{\m_j,\m_{j'}}(x_j, x_{j'}) \label{dl}
\ee
%
% using that
%
%\be
%G^\pm_{\m,i}(A+J)=e^{\pm e Q_i  \int_0^1 ds  J_m(x+t e_m a) } G^\pm_{\m,i}(A)
%+G^\pm_{\m,i}(J)
%\ee
%
%one gets 
%In order to perform the above integration we recall the simple expecttions
%
%\be
%rE_A(\prod_ {i=1}^n e^ {i \e_i \a_i A_{\m_i}(x_i)})=\d_{\sum_i \e_i,0} e^ {-{1\over 2}\sum_ {i,j}  \e_i \e_j \a_i \a_j g_{A,\m_i,\m_j}(x_i,x_j) }
%\ee
%
%which is equivalent to
%
%\be
 %E_A(\prod_ {i=1}^n e^{ {\a_i^2\over 2}  g_{A,\m_i,\m_i}(0,0)}}
%e^ {i \e_i \a_i A_{\m_i}(x_i)})=\d_{\sum_i \e_i,0} e^ {-\sum_ {i<j}  \e_i \e_j \a_i \a_j g_{A,\m_i,\m_j}(x_i,x_j) }
%\ee
%
The truncated expectation $\EE^T_A=e^{-V(X)}|_T$ are obtained by solving recursively
\be
e^{-V(X)}=\sum_\pi \prod_{Y\in \pi}e^{-V(Y)}|_T
\ee
where $\pi$ are the partitions of $X$. An explicit expression for the 
connected part is, (see \cite{Br} ) 
\be
e^{-V(X)}|_T=\sum_{g\in G} \prod_{(j,j')\in g}( e^{-V_{j,j'}}-1) \prod_{j\in X}e^{-V_{j,j}/2}\label{Br1}
\ee
where $G$ is the set of the connected graphs in $X$. 
A different representation for \pref{Br1} is however more convenient \cite{Br} (see also
\cite{Ben},\cite{BFM1})
, whose derivation is recalled in App. I
\be 
e^{-V(X)}|_T=\sum_{T\in  \TT }\prod_{(j,j')\in T} V_{j,j'}
\int dp_T(\underline s) e^{-V_T(\underline s)}\label{for}
\ee 
where 
\begin{itemize}
\item $\TT$ is the set of tree graphs $T$ on $X$
\item $\underline s\in (0,1)^{{\cal P}_X}$ with ${\cal P}_X$ the set of unordered pairs in $X$
\item $V_T(\underline s)$ is a convex linear combination of $V(Y)=\sum_{j,j'\in Y}V_{j,j'}$, $Y$ subsets of $X$.
\item $dp_T(\underline s)$ is a probability measure 
\end{itemize}
Note that $V(Y)$ is stable, that is
\be
V(Y)=\sum_{j,j'\in Y} V_{j,j}=\sum_{j,j'\in Y}\e^2 \b_j \b_{j'} 
a^2  g^A_{\m_i,\m_j}(x_i, x_j) =
\EE( [\sum_{i\in Y} \e \b_j  a  A_{\m_i}(x_i)]^2)\ge 0
\ee
hence 
\be
V(\underline s)\ge 0
\ee
and 
\be
\int dp_T(\underline s) e^{- V(\underline s)}\le 1
\ee
We can therefore write
\be
a^{-n}  |\EE^T_A(  e^{i\b_1 \e  a  
A_{\m_1} (x_1)} ;...;e^{i\b_n \e Q_n a
A_{\m_n
}
(x_n) })|\le C^n  \e^n a^{-n} 
 \sum_{T\in  \TT } \prod_{(i,j)\in T} 
|a^2  g^A_{\m_i,\m_j}(x_i, x_j)|
\ee
Using that
$|g^A_{\m,\m}|_1\le C M^{-2 }$
we get
\be 
a^{-n} 
 \sum_{T\in  \TT } \prod_{(i,j)\in T} |a^2 g^A(x_i,x_j)|_1\le \sum_{T\in  \TT } 
C^n \e^n a^{-n}  ({a^2\over M^2})^{n-1}\le
C^n \e^n n! a^{-n} a^{4n-4}  ({1\over a^2 M^2})^{n-1}
\ee
where we have used that the number of $T$ is $\le C^n n!$. Finally 
we can expand $e^{i\a_i a  J_{\m_i} }$
in series obtaining an extra $a$ for any $J$. 
\qed

\section{Integration of the fermionic field}

\subsection{Multiscale decomposition}

The {\it fermionic simple expectation} 
\be
\EE_\psi(\psi^{\e_1}_{i_1,x_1}...\psi^{\e_n}_{i_n,x_n})=\int P(d\psi) \psi^{\e_1}_{i_1,x_1}...\psi^{\e_n}_{i_n,x_n}
\ee
is expressed by the anticommutative Wick rule with covariance, $k={2\pi\over L}(n+{1\over 2})$, $n\in\NN$, $k\in [-\pi/a,\pi/a]^4$
\be
 g_i(x,y)=\int P(d\psi) \psi^-_{i, x} \psi^+_{i, y}={1\over L^4}\sum_k e^{i k(x-y)}
(\sum_{\m}i \g_0\g_\m a^{-1}  \sin (k_\m a)+ r  a^{-1}\g_0 \sum_\m (1-\cos k_\m a))^{-1}\label{propp}
\ee
The truncated expectations are $\EE^T_\psi(O;...;O)={\partial^n\over \partial \l^n} \log \int P(d\psi) e^{\l O}|_0$.
%For definiteness we start considering the case $\phi=0$ so that the generating function has the %following form
%
%\be
%e^{W(J,J^5, J^W),0}=\int P(d\psi) e^{V(\psi, J, j^A,j^W)}
%\ee
%
%with
%
%\be
%V=V_{A}
 %(\psi,J^{e.m.})+V_c(\psi)+V_Z(\psi) +
%\sum_i j^A_{x,i} \tilde J_{x,i}+\sum_{l=1}^3 Z^W_j j_{l,\m, x} J^W_\m}
%\ee
%The multiscale integration is defined inductively starting from the following decomposition of the %propagator . 

In contrast with the integration of the $A$ fields, the $\psi$ fields are massless and a multiscale integration procedure is necessary. If $\chi_0(t)$  is a Gevray function which is $=1$ for $t\le 1$ and $=0$ for $t\ge \g$ with 
$\g>1$, we define
\be
1=\chi_{N}(k)+ f_{N+1}(k)
\ee
where $\chi_{N}(k)=\sum_{h=-\io}^N f_h(k)$ with $f_h(k)=\chi_0(\g^{-h}  |k| )-\chi_0(\g^{-h+1}  |k| )$
and $f_h(k)$ non-vanishing for $\g^{h-1}\le |k|\le \g^{h+1}$; therefore
$\chi_{N}(k)=0$ for $|k|\ge \g^{N+1}=  \pi/4 a$ then $f_{N+1}(k)$ has support for $|k|\ge \pi/4 a$.
We define $\hat g^{(N+1)}_{i}(k)=f_{N+1}(k) \hat g_i(k)$; by \pref{propp}
\be
a^{-2}\sum_\m (\sin k_\m a)^{2}+ a^{-2}\sum_\m (1-\cos k_\m a)^{2}\ge 
a^{-2}\sum_\m (1-\cos k_\m a))^{2}\ge
C/a^2
\ee
for $|k|\ge \pi/4 a$, and using that the volume of the support is $O(a^{-4})$,
we get
\be
 |g^{(N+1)}(x,y)|\le C \g^{3 N} e^{-(c\g^N |x-y|)^{1\over 2}}
\ee
Regarding $g^{(h)}(x,y)$ in the support of $f^h$
one has that 
\be
a^{-2}\sum_\m (\sin k_\m a)^{2}+ a^{-2}\sum_\m (1-\cos k_\m a))^{2}\ge 
a^{-2}\sum_\m (\sin k_\m a)^{2}\ge
C \g^{2 h}
\ee
so that
\be
 |g^{(h+1)}(x,y)|\le C \g^{3 h} e^{-(c\g^h |x-y|)^{1\over 2}}
\ee
%
%we set $f_{h}(k)=\bar\chi (\g^{-h}( k )-\bar\chi (\g^{-h+1}( k )$
%and $\chi_{\le N-1}(k)=\sum_{h=-\io}^{N-1} f_h(k)$.
%By defining
%
%\be
%g^{(h)}(x)=\int dk f_h(k) e^{i k x}  \hat g(k)
%\ee
%
%Note that
%
%\be
%|g^h((k)|\le C \g^{3 h}  e^{-(\g^h |x|)^{1\over 2}}
%\ee
%
%In addition we can write, for $h\le N-1$
%
%\be
%g^h((k)={1\over L^4}\sum_k e^{i k(x-y)} {f_h(k)\over \g_0\g_\m _\m}+r^h(x)
%\ee
%
%with
%
%\be
%|r^h(x)|\le C \g^{4 h}  e^{-(\g^h |x|)^{1\over 2}}
%\ee
%
%which says that the single scale propagator coincides with the one of massless Dirac
%fermions up to smaller corrections
Finally
\be
g(x,y)=\sum_{h=-\io}^{N+1} g^{(h)}(x,y)
\ee
The multiscale integration is defined inductively in the following way; assume that we have integrated the fileds
$\psi^{(N+1)} , \psi^{(N-1)},... \psi^{(h)}$ obtaining (in the $\phi=0$ for definiteness)
\be
e^{W(J,J^5,0)}=\int P_{Z_h}  (d\psi^{(\le h)}) e^{V^{(h)}(\sqrt{Z_h}
\psi^{(\le h)},J,J^5)}
\ee
with $P(d\psi^{(\le h)})$ given by 
\be
\hat g_{i}^{(\le h)}(k)=\chi_h(k)
(\sum_{\m} \g_0\tilde\g^h_\m a^{-1}  i \sin (k_\m a)+  a^{-1}\hat\g^h_0 \sum_\m (1-\cos k_\m a))^{-1}\label{sapa}
\ee
\bea 
&&\tilde \g^h_0= \begin{pmatrix} 0& Z_{h,L,i} I  \\  Z_{h,R,i}I & 0  \end{pmatrix}\quad \tilde \g^h_j= \begin{pmatrix} 0& i Z_{h,L,i} \s_j \\-i Z_{h,R,i}\s_j& 0 \end{pmatrix}\nn\\
&& \hat \g^h_0= \begin{pmatrix} 0& r_h \sqrt {Z_{h,L,i}Z_{h,R,i} }   
I  \\  r_h\sqrt {Z_{h,L,i}Z_{h,R,i} }  I  & 0  \end{pmatrix}
\eea
and
\be
V^{(h)}(\psi^{(\le h)},J,J^5)=\sum_{l,m\ge 0\atop l+m\ge 1}\sum_{\underline s, \underline i,\underline \a, \underline \m}   \int d\underline x  d\underline y  
[\prod_{j=1}^l \partial^{\a_j}
\psi^{{(\le h)}\e_j}_{i_j ,s_j, x_j}] [\prod_{j=1} ^m  
J^{\s_j}_{\m_i,x_i} ]
W^{h}_{l,m,\underline \m, \underline s, \underline i,\underline \a} (\underline x, \underline y)
\ee
with $\s_j=0,1$ and $J^{0}_{\m,x}=
J_{\m}$, $J^{1}_{\m,x}=J^5_{\m,x}$; in $W^{h}_{l,m,\underline \m, \underline s, \underline i,\underline \a}$.
We define a localization operator $\LL$ such that 
\bea
&&\LL \int dx dy  W^h_{2,0}(x,y) \psi^{+(\le h)}_ {i,s,x} \psi^{-(\le h)}_ {i',s',y}=
\int dx dy W^h_{2,0}(x,y)\psi^{+(\le h)}_ {i,s,x} (\psi^{-(\le h)}_ {i',s',x}+
(x-y)_\m\tilde\partial_\m\psi^{-,(\le h)}_ {i',s',x})\nn\\
&&\LL \int dx dy dz W^h_{2,1,\m}(x,y,z)
J_{\m,z}^\s  \psi^{+(\le h)}_ {i,s,x}  \psi^{-(\le h)}_ {i',s',y}=\int dx dy dz
W^h_{2,1,\m}(x,y,z)
J_{\m,z}^\s  \psi^{+(\le h)}_ {i,s,z} \psi^{-(\le h)}_ {i',s',z}
\eea
where $\tilde\partial_\m f_x={f_{x+e_\m a}-f_{x-e_\m a} \over 2 a }$; $\LL=0$ otherwise.
In momentum space the localization operator can be written as  
\bea
&&\LL \int d k \hat W^h_{2,0}(k) \hat\psi^{+(\le h)}_{i,s,k} \hat\psi^{-(\le h)}_ {i',s',k}=
\int dk (\hat W_{2,0}(0)+i{\sin k_\m a\over a}\partial_\m \hat W_{2,0}(0))
\hat\psi^{+(\le h)}_{i,s,k} \hat\psi^{-(\le h)}_ {i',s',k}\\
&&\LL \int dk dp \hat W^h_{2,1,\m}(k,k+p)
\hat J_{\m,p}^\s  \hat\psi^{+(\le h)}_ {i,s,k} \hat \psi^{-(\le h)}_ {i',s',k+p}=\int dk dp \hat W^h_{2,1,\m}(0,0)
J_{\m,p}^\s  \hat\psi^{+(\le h)}_ {i,s,k} \hat \psi^{-(\le h)}_ {i',s',k+p} \nn
\eea
The action of $\LL$ gives 
\bea
&&\LL\VV^{(h)}(\sqrt{Z_h}\psi,J,J^5)
=\int dx \sum_{i,s}  n_{h,i}  \sqrt{Z_{h,i,L}Z_{h,i,R}}
 \g^{h}
(\psi^+_{i,L,x}\psi^-_{i,R,x}+\psi^+_{i,R,x}\psi^-_{i,L,x})+\nn\\
&&\sum_{i,s}  z_{h,i,s}Z_{h,i,s}\int dx  
\s_\m^s\psi^+_{i,s,x}\tilde\partial_\m\psi^+_{i,s,x}+
\nn\\ 
&&+
\sum_{i,s}   Z_{h,i,s}^{J} Y_i
\int dx 
J_{\m,x} \psi^+_{i,s,x}\s_\m^s \psi^-_{i,s,x}+\sum_{i,s} \tilde \e_i \e_s Y_i
 \int Z_{h,i,s}^{5} J^5_{\m,x}
\psi^+_{i,s,x}\s_\m^s \psi^-_{i,s,x}\label{ess11a}
\eea

Note that the local part of the electromagnetic and axial current, see the last line of 
\pref{ess11a}, is equal (up to a sign), even if the corresponding expression at scale $N$ were different (the vector current
is non local while the axial current is local).
%In order to write the above expression we have used that he system invariant under the global 
%transfornatioins
%
%\be
%\psi^+_{i,c, s,j,x}\to \psi^\pm_{i,c, s,j,x} e^{\pm i e Q_i \a}  \quad Q_u-Q_d=Q_e-Q_\n
%\quad J^\pm_W  \to   J^\pm_W e^{\pm i \a}
%label{as} u
%\ee
%
%with $J, J^5$ unchanged and 
%with $Q_i$ all different.  
Some symmetry considerations restrict the possible terms obtained by the $\LL$ operation.
By the invariance under $\psi^\pm_{i,s}\to e^{\pm i \a_i} \psi^\pm_{i,s}$
with $\a_\n-\a_\e=\a_u-\a_d$ and $\a_{i}\not =\a_{i'}$ we get
$\hat \psi^+_{i,s,k} \hat \psi^-_{i',s',k}\to
e^{i(\a_i-\a_{i'})} \hat \psi^-_{i,s,k} \hat \psi^+_{i',s',k}$ 
so that
$i=i'$.
%$W_{2,0}\psi^+\psi^-$,
%$W_{2,1} J\psi^+\psi^-$, $W_{2,1} J^5\psi^+\psi^-$ terms. Regarding 
%$J_W^+ \hat \psi^+_{i,s,k} \hat \psi^+_{i',s',k}$ the symmetry imposes $(i,i')=(e,\n); (e,d); (u,d);(u,\n)$,
%but no terms mixing different families are present by the symmetry \pref{cond3}.
%; one would need a odd number of term $j^W_q j^W_l$ leving some q fields uncontracted.
Regarding the chiral indices in  $W_{2,0}$
if $s=s'$ then 
$\hat W^h_{2,0}(k)$  is odd and the kernel with opposite chirality is even.
This says that in the $\hat W^h_{2,0}(0)$ terms the fields have opposite chirality and
$\partial_\m \hat W^h_{2,0}(0)$ the same chirality; for the same reason in $\hat W^h_{2,1}(0)$ they have the same chirality.
Finally
$\partial_\m W^h_{2,0}(0),W^h_{\m,2,1}(0) $ are proportional to $\s_\m^s$.

\subsection{Anomalous integration}
We can write 
\be
\tilde\LL\VV^{(h)}(\psi,J,J^5)=\LL\VV^{(h)}+\sum_{i,s}  z_{h,i,s} \int dx 
\s_\m^s\psi^+_{i,s,x}\tilde\partial_\m\psi^+_{i,s,x}\ee
so that
\be
e^{W(J,J^5,0)}=\int P_{Z_{h-1}}  (d\psi^{(\le h)}) 
e^{\tilde \LL V^{(h)}(\sqrt{Z_h}\psi^{(\le h)}, J,J^5) +\RR  V^{(h)}(\sqrt{Z_h}\psi^{(\le h)},J,J^5)}
\ee
where $P_{Z_{h-1}}  (d\psi^{(\le h)})$ has propagator given by \pref{sapa} with $Z_{h,i,s}$ replaced by
\be
Z_{h-1,i,s}(k)=Z_{h,i,s}+\chi_h(k) z_{h,i,s} 
\ee
Setting $Z_{h-1,i,s}(0)\equiv Z_{h-1,i,s}$, we can write
\be
P_{Z_{h-1}}  (d\psi^{(\le h)})= P_{Z_{h-1}}  (d\psi^{(\le h-1)})
P_{Z_{h-1}}  (d\psi^{(h)})
\ee
with $P_{Z_{h-1}}  (d\psi^{(\le h-1)})$ with propagator \pref{sapa} with $h$ replaced by $h-1$ and $P_{Z_{h-1}}  (d\psi^{(h)})$
has propagator which can be written as
\be
g^{(h)}(x,y)=g^{(h)}_{rel}(x,y)+r^{(h)}(x,y)\label{allo}
\ee
with
\be
g^{(h)}_{rel}(x,y)={1\over L^4} \sum_k e^{i k(x-y)} f_h(k) \begin{pmatrix} Z_{h-1,i,L}(k) (i k_0 I+\sum_j \s_j k_j)  & 0 \\   0& Z_{h-1,i,R}(k) (i k_0 I-\sum_j \s_j k_j)]
 \end{pmatrix}^{-1} 
\ee
and
\be
|g^{(h)}_{rel}(x,y)|\le 
C {\g^{3 h}\over Z_{h-1,i,L}}
e^{-(\g^h |x-y|)^{1\over 2}}\quad 
|r^{(h)}_{i} (x,y)|\le C \g^{3 h} \g^{h-N} e^{-(\g^h |x-y|)^{1\over 2}}
\ee
Finally we define
\be
e^{V^{(h-1)}(\sqrt{Z_{h-1}}\psi^{(\le h-1)},J,J^5)}=
\int P_{Z_{h-1}}  (d\psi^{(h)})e^{\tilde\LL V^{(h)}(\sqrt{Z_{h-1}}\psi^{(\le h)},J,J^5)+\RR  V^{(h)}(\sqrt{Z_{h-1}}\psi^{(\le h) },J,J^5)}
\ee
with
\bea
&&\tilde \LL\VV^{(h)}(\sqrt{Z_{h-1}}\psi,J,J^5)
=\int dx \sum_{i,s}  \n_{h,i}\sqrt{Z_{h-1,i,L}Z_{h-1,i,L}}
 \g^{h}
(\psi^+_{i,L,x}\psi^-_{i,R,x}+\psi^+_{i,R,x}\psi^-_{i,L,x})+\label{ann}\nn\\
%&&
%\sum_{j,j'=l,q}\sum_\m \sum_{s=L,R} Z^W_{h, k, j ,j' ,s,h-1}  \int dx  j^{k,W}_{\m,s,j',x}  J^{k,W}_{\m,j,x}  
%\nn\\ 
&&+
\sum_{i,s}   Z_{h-1,i,s}^{J} Y_i
\int dx 
J_{\m,x} \psi^+_{i,s,x}\s_\m^s \psi^-_{i,s,x}+
\sum_{i,s} \tilde \e_i \e_s Y_i
\int Z_{h-1,i,s}^{5} J^5_{\m,x}
\psi^+_{i,s,x}\s_\m^s \psi^-_{i,s,x}
\eea
and the procedure can be iterated. 

\subsection{Convergence and analyticity}

We prove the following lemma, setting $\n_{i,h}=\tilde\n_{i,h} ({a \bar M})^{-2}$.

\begin{lemma} 
There exists a constant $\bar\e$ such that,
for $|r_h|,  |Z_{h,i,s}|\le e^{C \bar\e ({a \bar M})^{2} }$,
$\max (|\tilde\n_{i,s,h}|,g^2,\bar g^2,\e^2 )\le \bar \e ({a \bar M})^{2} $  then
\be  
{1\over L^4}\int d\underline x d\underline y
|W^{(h)}_{l,m,\underline \m, \underline s, \underline i,\underline \a} (\underline x, \underline y)|
\le C^{l+m}  \g^{(4-(3/2)l-m) h} 
\bar\bar\e^{\max(l/2-1,1)} 
\label{bb}
\ee
\end{lemma}
\noindent
{\bf Proof} We write the kernels $W^{(h)}_{l,m,\underline \m, \underline s, \underline i,\underline \a}$ in in terms of {\it Gallavotti-Nicol\'o trees}  , see Fig.1, defined in the following way (for details see e.g.   \S 3 of \cite{Mbook}) .
\insertplot{300}{160}
{\ins{60pt}{90pt}{$v_0$}\ins{120pt}{100pt}{$v$}
\ins{100pt}{90pt}{$v'$}
\ins{120pt}{-5pt}{$h_v$}
\ins{235pt}{-5pt}{$N$}
\ins{255pt}{-5pt}{$N+1$}
}
{treelut2}{\label{n11} A labeled tree 
}{0}

%\insertplot{300}{200}
%{}
%{fig9a}{\label{n11} A graph with its clusters and the corresponding tree
%}{0}

Let us consider the family of all trees which can be constructed
by joining a point $r$, the {\it root}, with an ordered set of $n\ge 1$
points, the {\it endpoints} of the {\it unlabeled tree}, 
so that $r$ is not a branching point. $n$ will be called the
{\it order} of the unlabeled tree and the branching points will be called
the {\it non trivial vertices}.
The unlabeled trees are partially ordered from the root to the endpoints in
the natural way; we shall use the symbol $<$ to denote the partial order. 
The number of unlabeled trees is $4^n$.
The set of labeled trees $\TT_{h,n}$
is defined associating a label $h\le N-1$ with the root; 
moreover  
we introduce
a family of vertical lines, labeled by an integer taking values
in $[h,N+1]$ intersecting all the non-trivial vertices, the endpoints and other points called trivial vertices.
The set of the {\it
vertices} $v$ of $\t$ will be the union of the endpoints, the trivial vertices
and the non trivial vertices. The scale label is $h_v$ and, if $v_1$ and $v_2$ are two vertices and $v_1<v_2$, then
$h_{v_1}<h_{v_2}$.
Moreover, there is only one vertex immediately following
the root, which will be denoted $v_0$ and can not be an endpoint;
its scale is $h+1$.  

There are two kinds of end-points, normal and special,
and $n=\bar n+m$. The normal end-points are $\bar n$ and are associated to terms in the effective potential not depending on the external
fields $J$. The $\n$-end-points are associated with the first line of \pref{ann}
and have scale $h_v\le N+1$ and there is the constraint that  $h_v=h_{v'}+1$, if $v'$ is the first non trivial vertex immediately preceding $v$. The $V_F$-endpoints have scale $h_v=N+1$ and
are associated to one of the terms in \pref{quar}, and to the $V_A$-endpoints 
is associated one of the terms in \pref{sas3} with $m=0$.
The special end-endpoints have associated terms with at least an external $J$ fields;
the $Z^{J},Z^5$ end-points have $h_v\le N+1$ and there is the constraint that  $h_v=h_{v'}+1$, and are associated with one of the terms in
the second line of \pref{ann}; the $V_A$ end-points have scale $N$ and are associated to the terms 
with $(n,m)\not=(1,1)$ in \pref{sas3}.

%the scale is $N+1$ and are named as $\l$ or normal-end-points, or $A$ or special 
%endpoints $\BB(\psi^{(\le N)},A)$ or
%$\LL\VV^{h_v-1}(\psi^{(\le h_v-1)},A)$ and in this case the scale is $h_v \le N+1 $ 
%and there is the constraint that
%that $h_v=h_{v'}+1$, if $v'$ is the first non trivial vertex immediately preceding $v$;
%in such a case they are called special end-points.
%We will call $\bar m_{4,v}$ the number of $\l$ end-points with scale $h_v+1$,
%and $m_{4,v}=\sum_{\bar v\ge v}\bar m_{4,\bar v}$.
%To the special endpoints are associated the source terms linear in the external field $J$
%and have associated a rcc $Z_{h_v}$.
%Given $v\in\t$, we shall call
%$n^J_v$ the number of endpoints of type $J$ following $v$ in the tree.

The effective potential can be written as 
\be
\VV^{(h)}(\psi^{(\le h)},J,J^5) =
\sum_{n=1}^\io\sum_{\t\in\TT_{h,n}}
\VV^{(h)}(\t)\;,\ee
where, if $v_0$ is the first vertex of $\t$ 
and $\t_1,..,\t_s$ ($s=s_{v_0}$)
are the subtrees of $\t$ with root $v_0$,
$\VV^{(h)}$
is defined inductively by the relation, $h\le N-1$
$$\VV^{(h)}(\t)={(-1)^{s+1}\over s!} \EE^T_{h+1}[\bar
\VV^{(h+1)}(\t_1);..; \bar
\VV^{(h+1)}(\t_{s})]$$
where $\EE^T_{h+1}$ is the {\it truncated expectation}
and $\bar
\VV^{(h+1)}(\t)=\RR \VV^{(h+1)}(\t)$ if
the subtree $\t_i$ contains more then one end-point.
%while if $\t_i$ contains only one end-point $\bar
%\VV^{(h+1)}(\t)$ is
%$V(\psi^{(\le N)})$ if is a normal end-point (and in such case $h=N-1$)  
%or if is a special end-point
%$\LL\VV^{h+1}(\psi^{(\le h+1)},J,J^5,J^W)$, $h<N-1$ or $\BB(\psi^{(\le N)},J,J^5,J^W)$.
We define
$P_v$ as the set of field labels of $v$
representing the external fields 
and if
$v_1,\ldots,v_{s_v}$ are the $s_v$
vertices immediately following $v$, then we denote by $Q_{v_i}$ the intersection of $P_v$ and
$P_{v_i}$; this definition implies that $P_v=\cup_i Q_{v_i}$. The
union of the subsets $P_{v_i}\bs Q_{v_i}$ are the internal fields of $v$.
Therefore if $\PP_\t$ is the familiy of all such choices and $\bP$
an element we can write
\be
 \VV^{(h)}(\t)=\sum_{\bP\in\PP_\t}
\int dx_{v_0} W^{(h+1)}_{\t,\bP}(x_{v_0})[\prod_{f\in P_{v_0}}
\psi^{\e(f)(\le h)}_{x(f)}][\prod_f J(x_f)]
\ee
where $W^{(h_v)}_{\t,\bP}(x_{v_0})$ is defined inductively by the equation
\be
W^{(h+1)}_{\t,\bP}(x_{v})={1\over s_{v} !} [\prod_{i=1}^{s_{v_i}}W^{(h_v+1)}_{\t,\bP}(x_{v_i})] \EE^T_{h_v}(
\tilde\psi^{(h_v)}(P_{v_1}/Q_{v_1});
...;\tilde\psi^{(h)_v}(P_{v_{s_v} }/Q_{v_{s_v}})
\ee
where $\tilde\psi^{(h)}(P)=\prod_{f\in P}\psi^{(h)\e(f)}_{x(f)}$ and $x_v$ are the coordinates associated to the vertex $v$.

We get
\be
W^{(h)}=\sum_{\t\in \TT_{h,n}}\sum_{\bP, |P_{v_0}|=l+m}
W^{(h)}_{\t,\bP}(x_{v})
\ee
We use the analogue of \pref{for} for fermionic
truncated expectation,
\be
\EE^T_{h}(\tilde\psi^{(h)}(P_1);\tilde\psi^{(h)}(P_2);
...;\tilde\psi^{(h)}(P_s))= \sum_{T}\prod_{l\in T} g^{(h)}(x_l-y_l)
\int dP_{T}(\tt) \det
G^{h,T}(\tt)\label{xx}\ee
where
\begin{enumerate}
\item  
 $T$ is a
set of lines, which becomes a tree graph if one identifies all the
points in the same cluster.
\item  $\tt=\{t_{i,i'}\in [0,1],
1\le i,i' \le s\}$ and $dP_{T}(\tt)$ is a probability measure with
support on a set of $\tt$ such that $t_{i,i'}=\uu_i\cdot\uu_{i'}$
for some family of vectors $\uu_i\in \RRR^s$ of unit norm. 
\item 
$G^{h,T}(\tt)$ is a $(n-s+1)\times (n-s+1)$ matrix, whose elements
are given by
$G^{h,T}_{ij,i'j'}=t_{i,i'} 
g^{(h)}(x_{ij}-y_{i'j'})$.
\item If $\HH=\RRR^s\otimes \HH_0$, where $\HH_0$ is the Hilbert space of complex
two dimensional vectors with scalar product
$<F,G>=\int dk F^*_i(k) G_i(k)$.
It is easy to verify that
\be
G^{h_v,T_v}_{ij,i'j'}=t_{i,i'} g^{(h_v)}(x_{ij}-y_{i'j'})
=<\uu_i\otimes A^{(h_v)}_{x(f^-_{ij})
},
\uu_{i'}\otimes B^{(h_v)}_{x(f^+_{i'j'})}>\ee
where $\uu_i\in \RRR^s$, $i=1,\ldots,s$, are the vectors such that
$t_{i,i'}=\uu_i\cdot\uu_{i'}$ and $A, B$ suitable functions. 
\end{enumerate}

By inserting the above representation we can write
$
W^{(h+1)}_{\t,\bP}=\sum_T W^{(h+1)}_{\t,\bP,T}
$
where $\bT$ is the union of all the trees $T$.

The determinants are bounded by the {\it Gram-Hadamard inequality}, stating
that, if $M$ is a square matrix with elements $M_{ij}$ of the form
$M_{ij}=<A_i,B_j>$, where $A_i$, $B_j$ are vectors in a Hilbert space with
scalar product $<\cdot,\cdot>$, then
\be
|\det M|\le \prod_i ||A_i||\cdot ||B_i||\label{gh} 
\ee
where $||\cdot||$ is the norm induced by the scalar product.

The tree $T$ connects a set of $n$-endpoints; to each point $v$ are associated $i_v$ $\psi$-fields and $j_v$ $J$-fields;
we define $m^{i,j}_v$ the number of end-points following $v$  with $i$ $\psi$ fields and $j$ $J$ fields. 
The integral over the difference
of coordinates associated to propagators gives a factor
$\prod_v \g^{-4 h_v(s_v-1)}$; to the normal $V_F$ end-points is associated $\g^{-2N} (1/(\bar Ma)^2)$;
to the 
$V_A$ endpoints is associated
a factor $ \g^{(4-3 i_v/2-j_v)N} (\e^{i_v/2} (a \bar M)^{2-i_v})$
with $(4-3 i_v/2-j_v)<0$ and $i_v\ge 4$ and $(a \bar M)^{2-i_v}<(a \bar M)^{-2}$.
%(${1\over (\bar M a)^{2n-2}}\le  {c_1^{4-2n}  \over (\bar M a)^{2}} $ as $n\ge 2$) .
In conclusion we get
%, using 
%$\sum_{n=2}^\io  e^{n} (a \bar M)^{-2 n+2}\le C (e^2 (a \bar M)^{-2})$
% using that for any $K$
%$ \int dx |v(x)|\le  \g^{-2 N}$
%and
%
%\be
%|g^h(x)|\le C \g^{3 h} e^{-(\g^h |x|)^{1\over 2}} \quad\quad \int dx |g^h(x)|\le %C  \g^{- h} 
%\ee
%
%In conclusion we get
%
\bea
&&\int d x_{v_0} |W_{\t,\bP,T}(x_{v_0})|\le L^4 \prod_{v\, not\, e.p.} {1\over s_v!} 
C^{\sum_{i=1}^{s_v}|P_{v_i}|-|P_v|} \g^{-4 h_v (s_v-1)}\\
&&
\g^{3/2 h_v (\sum_i |P_{v_i}|-|P_v|)}[\prod_v \g^{-z_v(h_v-h_{v'}  ) } ]
[\prod_{v\;e.p. \; not\; \n}
 \g^{(4-3 i_v/2-j_v)N} ][\prod_{v  \;e.p.\;\n} \g^{h_v} ] 
\bar\e^{\bar n}\nn
\eea
where $\prod_{v\;e.p. \; not; \n}$ is the product over the end-points excluded the $\n$ ones,
$\bar n$ is the number of normal end-points,$ \prod_v \g^{-z_v(h_v-h_{v'}  ) }$ is produced by the $\RR$ operation and
$z_v=2$ if $|P_v|=2$ and there are no $J$ fields, $z_v=1$ if $|P_v|=2$ and there is a single $J$ field, $z_v=0$ otherwise.
Note that $\prod_{v\;e.p. \; not\; \n}
 \g^{(4-3 i_v/2-j_v)N}$ includes the contributions of the $Z$ special end-points, where $4-3 i_v/2-j_v=0$.
By using that 
\bea
&&\sum_v (h_v-h)(s_v-1)=\sum_v (h_v-h_{v'})(\sum_{i,j } m^{i,j}_v-1)\nn\\
&&\sum_v (h_v-h) (\sum_i |P_{v_i}|-|P_v|)=\sum_v (h_v-h_{v'})(\sum_{i,j} i  m^{i,j}_v-|P_v|)
\eea
where $m^{i,j}_v$ is the number of end-points following $v$  with $i$ $\psi$ fields and $j$ $J$ fields , we get, if $\bar n$ is the number of normal endpoints
\bea
&&\int dx_{v_0} |W_{\t,\bP,T}(x_{v_0})|\le L^4 
\g^{-h[-4+{3|P_{v_0}|\over 2}- \sum_{i,j} (3 i/2-4) m^{i,j}_{v_0}]} 
\bar\e^{\bar n}
\nn\\
&&\prod_{v\; not \; e.p.} \left\{ {1\over s_v!}
C^{\sum_{i=1}^{s_v}|P_{v_i}|-|P_v|}
\g^{-(-4+{3|P_v|\over 2}-\sum_{i,j}(3 i/2-4) m^{i,j}_v+z_v)
(h_v-h_{v'})}\right\}\nn\\
&& [
\prod_{v\;e.p. \; not\; \n}\g^{(4-3 i_v/2-j_v)N}][\prod_{v\; e.p.\; \n} \g^{h_v}]
\nn\eea
We use now that
\be
\g^{h \sum_{i,j } m^{i,j}_{v_0}}\prod_{v\; not \; e.p.} \g^{\sum_{i,j }(h_v-h_{v'}) m^{i,j}_v}=\prod_{v \; e.p.} \g^{h_{v^*}}
\ee
where $v^*$ is the first non trivial vertex following $v$; this implies
%
%where $\bar m_v$ is the number of end-points internal to $v$ and not to any smaller one; therefore
%
\be
\g^{h\sum_{i,j} (3 i/2-4) m^{i,j}_{v_0}}\prod_{v\; not \; e.p.} \g^{\sum_{i,j}(3 i/2-4) m^{i,j}_v(h_v-h_{v'}) }
=\prod_{v\; e.p\; not\; \n} \g^{h_{v^*} (3 i_{v}/2-4)}\prod_{v \;  e.p.\; \n} \g^{-h_v}
\ee
so that
\bea
&&\int dx_{v_0} |W_{\t,\bP,T}(x_{v_0})|\le L^4 
\g^{-h[-4+{3|P_{v_0}|\over 2}]} 
\bar\e^{\bar n} 
\prod_{v\; e.p\; not\; \n} \g^{h_{v^*} (3 i_{v}/2-4)} \e^{\bar n} \nn\\
&&\prod_{v\; not\; e.p.} \left\{ {1\over s_v!}
C^{\sum_{i=1}^{s_v}|P_{v_i}|-|P_v|}
\g^{-(-4+{3|P_v|\over 2}+z_v)
(h_v-h_{v'})}\right\} [
\prod_{v\;e.p. \; not\; \n}\g^{(4-3 i_v/2-j_v)N}]
\eea
Finally we use the relation
\be
[\prod_{v\; e.p.} \g^{h_{v^*} j_v}][ \prod_{v \; e.p. } \g^{-h_{v^*} j_v}]=[\prod_{v\; e.p.} \g^{h_{v^*} j_v}]
\g^{-h \sum_{i,j } j m^{i,j}_{v_0}}\prod_{v\; not \; e.p.} \g^{-\sum_{i,j }(h_v-h_{v'}) j m^{i,j}_v}
\ee
and using that $\sum_{i,j}   j m^{i,j}_v  =n^J_v$ we finally get  ($j_v=0$ if $v$ is a $\n$-e.p.) 
\bea
&&\int dx_{v_0} |W_{\t,\bP,T}(x_{v_0})|\le L^4 
\g^{-h[-4+{3|P_{v_0}|\over 2}+n^J_{v_0}  ]} 
\bar\e^{\bar n}
\nn\\
&&\prod_{v\,\hbox{\ottorm not e.p.}} \left\{ {1\over s_v!}
C^{\sum_{i=1}^{s_v}|P_{v_i}|-|P_v|}
\g^{-(-4+{3|P_v|\over 2}+z_v+n^J_v )
(h_v-h_{v'})}\right\} [
\prod_{v\;e.p. \; not\; \n}\g^{(4-3 i_v/2-j_v)(N-h_{v^*})}]
\nn\eea
In conclusion
\bea
&&\int d x_{v_0} |W_{\t,\bP,T}(\xx_{v_0})|\le L^4 
\g^{-h d_{v_0}} C^n \bar\e^{\bar n}\nn
\\
&&[\prod_{\tilde v} 
{1\over s_{\tilde v}!}\g^{-d_{\tilde v} (h_{\tilde v}-h_{\tilde v'}) }
][
\prod_{v\;e.p. \; not; \n}\g^{(4-3 i_v/2-j_v)(N-h_{v^*})}]\label{ess}
\eea
where: $\tilde v\in \tilde V$ are the vertices on the tree such that $\sum_i |P_{v_i}|-|P_v|\not=0$, $\tilde v'$ is the
vertex in $\tilde V$ immediately preceding $\tilde v$ or the root;
$d_v=-4+{3|P_v|\over 2}+n^J_v+z_v$.
Finally the number of addenda in $\sum_{T\in {\bf T}}$ is bounded by
$\prod_{v} s_v!\;
C^{\sum_{i=1}^{s_v}|P_{v_i}|-|P_v|}$. 
In order to bound the sums over the scale labels and $\bP$ we first use
the inequality
\be
\prod_{\tilde v} \g^{-d_{\tilde v} (h_{\tilde v}-h_{\tilde v'}) }
\le [\prod_{\tilde v} \g^{-{1\over 2}(h_{\tilde v}-h_{\tilde v'})}]
[\prod_{\tilde v}\g^{-{3|P_{\tilde v}|\over 4}}]
\ee
where $\tilde v$ are the non trivial vertices, and $\tilde v'$ is the
non trivial vertex immediately preceding $\tilde v$ or the root. The
factors $\g^{-{1\over 2}(h_{\tilde v}-h_{\tilde v'})}$ in the r.h.s.
allow to bound the sums over the scale labels by $C^n$.
\qed
\vskip.3cm
An immediate corollary of the above proof is the following.
\begin{lemma} If $\TT^*$ is the set of trees with at least an end-point not of $\n, Z$ type then, for $0<\th<1$
\be
\sum_{\t\in \TT^*}
\sum_{\bP,T}
\int d x_{v_0} |W_{\t,\bP,T}(x_{v_0})|\le L^4 \g^{(4-(3/2)l-m) h} \g^{\th (h-N)}
\bar\e^{\max(l/2-1,1)} \label{lus}
\ee
\end{lemma}
\noindent
{\bf Proof} Let be $\hat v$ the non trivial vertex following an end-point
not of $\n, Z$ type; hence we can rewrite in \pref{ess} 
\be
[\prod_{\tilde v}
\g^{-d_{\tilde v} (h_{\tilde v}-h_{\tilde v'}) }] =
[\prod_{\tilde v}
\g^{-(d_{\tilde v}-\th) (h_{\tilde v}-h_{\tilde v'}) }] \g^{\th (h-h_{\hat v}) }
\ee
and 
\be
\g^{\th (h-h_{\hat v}) }[
\prod_{v\;e.p. \; not\; \n}\g^{(4-3 i_v/2-j_v)(N-h_{v^*})}]\le  \g^{\th (h-N) }
\ee
as $\prod_{v\;e.p. \; not\; \n}\g^{(4-3 i_v/2-j_v)(N-h_{v^*})}\le  \g^{-\th (N-h_{\hat v}) }$ as there is at least an end-point not $\n,Z$.
Noting that $d_{\tilde v}-\th>0$ one can perform the sum as in Lemma 4.1, and the same bound is obtained with an extra
$\g^{\th (h-N)}$.
\qed

%The existence of the limit $L\to\io$ is an immediate consequence of the tree expansion, see e g \cite{GM}.
%Analyticity stated  in
%is an immediate consequence of the prrof. Note that the denominator of the correlations (the partition function)
%at finite $L$ is analytic for any $\l$ in the whole complex plane as it is a finite dimensional Grasmann integral; on the other hand the RG analysis above provides an %expansion
%which coincides order by order and is analytic in a finite domain, so that it fully reconstructs the partition function.
%The correlation is also analytic, as the denominator is non vanishing in a finite %disk for small $\l$
%for any $L$
%and the numerator is a finite dimensional integral; moreover
%it coincides order by order with the expansion found analyzing the generating function by RG
%which is also analytic in the same domain so that they coincide and analyticity as $L\to\io$ follow

%Note that if we replace $e^2$ with $\l$ in $V_{ij}$ in \pref{dl}, the above %results implies analitycity in $\l$.

\section{Running coupling constants}

%The factor $\g^{ \th (h-N)}$ is a gain with respect to the "dimensional bound" in the term with at %least a $\l$, and is due to the dimensional irrelevance of the quartic terms; such extra factor plays a %crucial role in the following.
%Note that the estimated convergence radius is proportional to the cut-off and mass ratio, as a %consequence of the perturbative non-renormalizability of the theory.

It is a byproduct of the tree expansion defined in the previous section that $\n_{h,i}$ 
%$\ZZ_h=(Z_{h,i,s},Z^5_{h,i,s},
%Z^W_{h,i,s})$ 
verifies the following equation
\be
\n_{h-1,i}=\g \n_{h,i}+\b^{(h)}_{\n,i}\label{tok}
\ee
where
\be
\b^{(h)}_{\n,i}=\sum_{n=2}^\io \sum_{\t\in \TT^*_n}
\sum_{\bP,T} {1\over L^4}\int d x_{v_0} |W_{\t,\bP,T}(\xx_{v_0})|
\ee
where $|P_{v_0}|=2$, $h_{v_0}=h+1$ is a non trivial vertex and $\TT^*$ 
is the set of trees with at least a normal end-points not of $\n,Z$ type; this last condition is due to the fact that
in momentum space the kernels are computed at vanishing momenta and when
only end-points $\n$ are present only chain graphs contribute; therefore they are vanishing as $\hat g^{(h)}(0)=0$
by the compact support properties of the single scale propagator. Therefore by \pref{lus}, if $\tilde g^2=\max (g^2,\bar g^2, \e^2)$
\be
|\b^{(h)}_{\n,i}|\le C \g^{\th (h-N)}
[\max (\tilde g^2,\tilde\n_h)(M a)^{-2}]^{2}
\ee
The tree expansion is convergent provided that $(g^2,\bar g^2, \e^2,\tilde\n_h)(M a)^{-2}$ is smaller than some constant; 
this condition can be verified choosing 
$\tilde g^2\le c (M a)^2$, a condition which can be always verified. One needs however a similar condition on $\n_h$,
which according to \pref{tok} is generically $O(\g^{-h})$; it is possible however to suitably choose $\n\equiv \n_N$ so that $\n_h$ is bounded for any $h$.
We can rewrite \pref{tok}  as
\be
\n_{h-1,i}=\g^{-h}(\g^N\n_{i}+\sum_{k=h}^N \g^k \b^{(k)}_{\n,i})\label{tok}
\ee
We consider the system, by imposing the condition $\n_{-\io,i}=0$
\be
\n_{h-1,i}=\g^{-h}(-\sum_{k\le h} \g^k \b^{(k)}_{\n,i})\label{mklh}
\ee
Note that $\b^{(h)}_{\n,i}$ is a function of $g,\bar g, e$ and of the effective parameters $\n_{i,k}$ with $k\ge h$. Therefore,
we can regard the right side of \pref{mklh}
as a function of the whole sequence $\n_{i,k}$, which we can denote by
$\underline \n=\{\n_k\}_{k\le N}$
so that \pref{mklh}
can be read as a fixed point equation $\underline \n=T(\underline \n)$
on the Banach space of sequences $\n$ 
such that $||\n||=\sup_{k\le N} \g^{\th (k-N)} |\n_k|\le C \tilde g^2 (M a)^{-2}
$. It is a corollary of the proof in Lemma 4.1 (see e.g.  App A5 of \cite{GM1} for details)
that 
there is a choice of $\n_i$ such that the sequence is bounded for any $h$.
Therefore for a proper $\n$
\be
|\n_k|\le C   \g^{\th (h-N)}  \tilde g^2 (M a)^{-2}\label{bb1}
\ee
It remains to check the condition on boundedness of the effective renormalizations.
They verify recursive equations, if $\ZZ_h=(Z_{h,i,s},Z^{J}_{h,i,s }, Z^5_{h,i,s})$
\be
\ZZ_{h-1}=\ZZ_h+\b^{(h)}_\ZZ(\ZZ_h,..,\ZZ_N)\quad\quad |\b^{(h)}_\ZZ|\le C \g^{\th (h-N)} (g^2,\bar g^2, \e^2)(M a)^{-2}
\label{gamma1}
\ee
from which 
\be
\ZZ_{h-1}=\ZZ_N+\sum_{k=h}^N \b^{(k)}_\ZZ(\l;\ZZ_k,..,\ZZ_N)\label{sep} 
\ee
Therefore 
%
%\be
%|\ZZ_{-\io}-\ZZ_N|\le C 
% \tilde g^2 (M a)^{-2}
%\label{sep} 
%\ee
%
%and
%
\be
|\ZZ_{-\io}-\ZZ_h|\le C \g^{\th (h-N)} 
\tilde g^2 
(M a)^{-2}
\label{sep11} 
\ee
By the above bound it follows that we can choose
%$Z^{e.m}_{i,i',s,s'}=O(\tilde g^2(M a)^{-2})$ with $i\not i'$ or $s\not s'$ so that 
%$Z^{e.m}_{-\io i,i',s,s'}=0$; similarly we choose 
$Z^5_{i,s}$ so that
\be
Z^5_{-\io,i,s}=Z^{J}_{-\io,i,s}\label{sep12}
\ee
The existence of the limit $L\to\io$ for the kernels and expectations 
is an immediate consequence of the tree expansion; for an explicit derivation see e.g. App. D of \cite{GMaa}.
Analyticity of the correlations 
is an immediate consequence of the proof. Note that the denominator of the correlations (the partition function)
at finite $L$ is analytic in the whole complex plane as it is a finite dimensional Grasmann integral; on the other hand the RG analysis above provides an expansion
which coincides order by order and is analytic in a finite domain, so that it fully reconstructs the partition function.
The correlation is also analytic, as the denominator is non vanishing in a finite disk for small $\e^2,g^2, \bar g^2$
for any $L$
and the numerator is a finite dimensional integral; 
it coincides order by order with the expansion found analyzing the generating function by RG
which is also analytic in the same domain so that they coincide.

\section{Proof of Theorem 1.1}

In order to compute the 2-point function we have to consider the generating function with $\phi\not=0$; by a straightforward adaptation of the tree expansion,
(for details see e.g. \S 3.D of \cite{GMaa}),  one gets
\be
\hat S_i(k)=
(\sum_{\m} \tilde \g^{-\io}_\m a^{-1}  \sin (k_\m a)+  a^{-1}\g_0 \sum_\m (1-\cos k_\m a))^{-1}(I+R(k))\label{all1}
\ee
with
\be \tilde \g^{-\io}_0= \begin{pmatrix} Z_{-\io,i,L} I & 0 \\   0& Z_{-\io,i,R} I  \end{pmatrix}\quad \tilde \g^{-\io}_j= \begin{pmatrix} i Z_{-\io,i,L} \s_j & 0
\\0& -i Z_{-\io,i,R}\s_j \end{pmatrix}
\ee

and 
\be
|R(k)|\le C  \tilde g^2 |k a|^\th  
\ee
where the extra factor $|k a|^\th$ follows from \pref{lus}. 

Regarding the vertex function, we can again separate the contribution from trees involving only $Z$ vertex from the rest, which has by lemma 3.2 an improvement
in the bound $O( |a \k|^\th)$, with $\k=max(|k|,|k+p|)$, so that
\be\hat \G_{\m,i,s}(k,p)  = \hat g_{i,s,s} (k) \hat  g_{i,s,s}(k+p) 
[\s_\m^s {Z^{J}_{-\io, i,s}\over Z^2_{-\io,i,s}  } +O( |a \k|^\th)]\label{all}
%&&\hat \G^{W}_{\m,l,e,\n,s}(k,p)  = \hat g_{e,s,s} (k) \s_\m^L\hat  g_{\n,s,s}(k+p) 
%[{Z^{W}_{,s,-\io}\over Z_{\n,s,-\io} Z_{e,L,-\io} } +O(|a \k|^\th)]\nn\\
%&&\hat \G^{W}_{\m,q,u,d,s}(k,p)  = \hat g_{u,s,s} (k)\s_\m^L \hat  g_{d,s,s}(k+p) 
%[{Z^{W}_{q,s,-\io}\over Z_{u,s,-\io} Z^L_{d,s,-\io} } +O( |a \k|^\th)]\nn\\
%&&\hat \G^{5}_{\m,i,i,s}(k,p)  = \hat g_{i,s,s} (k)\s_\m^L \hat  g_{i,s,s}(k+p) 
%[ \e_s {Z^{5}_{i,s,-\io}\over  Z^2_{s,i,-\io} } +O( |a \k|^\th)]\nn\\
\ee
%Note that the lowest order contribution proportional to ${Z^{e.m.}_{i,s,i',s',-\io}\over Z^2_{s,i,-\io} }$ with $i\not=i'$
%or $s\not=s'$is vanishing. 
%
%By the above expressions we get, recalling \pref{al} and \pref{emp}
%\bea
%&& Z_{s,i,-\io}=Z^D_{s,i}\quad Z^{5}_{i,s,-\io}=Z^{e.m.}_{s,i,-\io}\quad\quad
%{Z^{W}_{l,L,-\io}\over \sqrt{Z^L_{\n,L,-\io} Z^L_{e,L,-\io}}}=
% {Z^{W}_{q,L,-\io}\over \sqrt{Z^L_{u,L,-\io} Z^L_{d,L,-\io}}}=1\nn\\
%&&{Z^{W}_{l,R,-\io}\over \sqrt{Z^R_{\n,L,-\io} Z^R_{e,L,-\io}}}=
%{Z^{W}_{q,R,-\io}\over \sqrt{Z^R_{u,L,-\io} Z^R_{d,L,-\io}}}=0
%\eea
%
We consider now the three current correlations defined above; it turns out that  
\be
\hat \Pi^5_{\m_1, \m_2,\m_3}(p_1,p_2)=\sum_{h=-\io}^N \int dx_2 dx_3
W^h_{0,3}(0,x_2,x_3) e^{-i p_1 x_2 -ip_2 x_3}
\ee
and using \pref{bb},\pref{bb1}, \pref{sep} 
we get
\be
|\hat \Pi^5_{\m_1,,\m_2,\m_3}(p_1,p_2)|\le \sum_{h=-\io}^N  C \g^h\le \tilde C
\ee	
with $\tilde C$ dependent on $N$.
The Fourier transform is therefore bounded and, in the limit $L\to\io$, continuous in $p_1,p_2$; it is however non differentiable.
We call $\TT_h^0$ the trees with only three special end-points $\ZZ_h$ and $\TT_h^1$ the set of the remaining trees, see Fig. 2 .
The contributions from $\TT_h^1$ to $\partial\hat \Pi_{\m_1,\m_2,\m_3}(p_1,p_2)$ have an extra $\g^{\th (h-N)}$, by Lemma 3.1
and \pref{bb1} and bounded by \be \sum_{h=-\io}^N
\tilde g^2 \g^{\th (h-N)}\le C \tilde g^2\ee
hence are differentiable.
We consider now the contribution from $\TT^0_h$.
We can write the propagator as \pref{allo};
the contribution with at least a propagator $r^{(h)}_{ i, s,s'}$ have an extra factor $\g^{h-N}$ and are therefore differentiable.
\insertplot{230}{89}
{\ins{50pt}{40pt}{$=$}
\ins{130pt}{40pt}{$+$}
}
{figjsp44a}
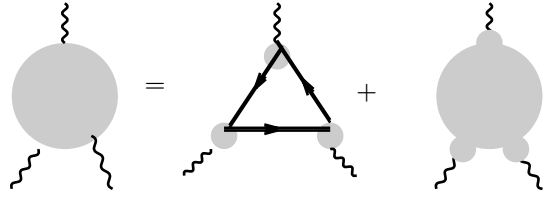
{\label{h2} Decomposition in terms of trees $\TT^0_h$ and $\TT^1_h$. 
} {0}
We consider now the terms belonging to $\TT^0_h$ with only propagators 
$g^{(h)}_{rel}(\xx,\yy)$; they are given by triangle graphs and 
to the axial
vertex is associated $Z^5_{h,i,s}$ and  to the vector vertex is
associated $Z^{J}_{h,i,s}$ (with the same $i,s$ as the index is conserved in the loop). 
We can replace the renormalizations $\ZZ_h$
with $\ZZ_{-\io}$ and use \pref{sep12}, \pref{wib}; 
the corrections has again an extra $\g^{\th(h-N)}$ by \pref{sep11} and are therefore differentiable.
We finally get
%is associated, by \pref{sep11}, 
%\pref{sep12} 
%, the sum of terms of the form
%\be
%{1\over 2}(Z^{e.m.}_{i,s,k}+ Z^{5}_{i,s,k}) \s_\m^s=  Z^{e.m.}_{i,s,k}
%\s_\m^s+R_{i,s,k} \quad |R_{i,s,k}|\le C \tilde g^2 \g^{\th(k-N)} (M a)^{-2}
%\ee
%
%by \pref{sep11},\pref{sep12}.
%
\be
 \hat \Pi^5_{\m,\r,\s}(p_1,p_2)=\hat \Pi^a_{\m,\r,\s}(p_1,p_2)+\hat\Pi^b_{\m,\r,\s}(p_1,p_2)\label{fon}
\ee
where $\Pi^b_{\m,\r,\s}$ is differentiable and
\bea &&\hat \Pi^a_{\m,\r,\s}(p_1,p_2)=
\sum_{h_1\atop h_2,h_3} \sum_{i,s} \tilde \e_i \e_s  Y_{i}^3 { Z^{5}_{-\io,i,s}
\over Z_{-\io,i,s}} {Z^{J}_{-\io, i,s}\over Z_{-\io,i,s}}
{Z^{J}_{-\io, i, s}\over Z_{-\io,i,s}}\nn\\
&&\int {dk \over (2\pi)^4}{\rm Tr}{f_ {h_1}(k)\over i  \s^{s}_\m k_\m}
i\s^{s}_\m{ f_{h_2}(k)\over i \s^{s}_\m (k_\m+p_\m)} i \s^{s}_\n 
{f_ {h_3}(k)\over i   \s^{s}_\m (k_\m+p^2_\m)}(i \s^{s}_\r)\label{ssas}
\eea
The renormalizations $\ZZ_{-\io}$ depend from the particle species and the chirality. They are however the same appearing in the 2-point and vertex correlations so that we can use the Ward Identities;
by inserting \pref{all1}, \pref{all} in \pref{wia} we get
\be {Z_{-\io,i,s}^{J}
\over Z_{-\io,i,s}}=1\ee
In conclusion 
\be
\hat\Pi^5_{\m,\r,\s}=\hat I_{\m,\r,\s}+\hat \RR_{\m,\r,\s}
\ee
with $\hat \RR$ with H\"older continuous derivative and 
\be
\hat I_{\m,\r,\s}(p_1,p_2)=
(\sum_i \tilde \e_i Y_i^3)
\int {dk \over (2\pi)^4}{\rm Tr} {\chi(k)\over \not k }\g_\m \g_5 {\chi(k+p)\over \not k+
\not p}\g_\n{\chi(k+p^2)\over \not k+\not p^2}\g_\s
\ee
Note that $\hat I$ is the anomaly for relativistic continuum fermions 
with a momentum regularization which
violates the vector current conservation \cite{BGM1}, \S 3.6 
\be 
\sum_\m (p_{1,\mu} + p_{2,\mu}) \hat I_{\mu,\nu,\sigma}
 = {(\sum_i \tilde \e_i Y_i^3)\over 6\pi^{2}} p_{1,\alpha} p_{2,\beta} \varepsilon_{\alpha\beta\nu\sigma}\quad
\sum_\n p_{1,\nu} \hat I_{\mu,\nu,\sigma}={(\sum_i \tilde \e_i Y_i^3)\over 6\pi^{2}} p_{1,\alpha} p_{2,\beta} \varepsilon_{\alpha\beta\mu\sigma}
\ee
 up to 
$O(a^\th |\bar p|^{2+\th})$ corrections. 
In contrast with $\hat I_{\m,\r,\s}$, we have that $\hat\RR_{\m,\r,\s}$ has not a simple explicit expression, being expressed in terms of a convergent series depending on all the lattice and interaction details. We use the differentiability of 
$\hat\RR_{\m,\r,\s}(p_1,p_2)$ to expand it at first order
obtaining, again up to  
$O(a^\th |\bar p|^{2+\th})$ corrections, using the current conservation \pref{wib} 
%$
%\hat \Pi_{\m,\n,\s}(p_1,p_2)=\hat I_{\m,\r,\s}(p_1,p_2)$
%\be
%+\hat\RR_{\m,\n,\s}(0,0)+\sum_{\rho} \sum_{a=1,2}  p_{a,\rho} 
%\frac{\partial \hat\RR_{\mu,\nu,\sigma}}{\partial p_{a,\rho}}({0}, {0}) 
%label{fon1}
%\ee
%
%
\be
\frac{1}{6\pi^{2}}
(\sum_i \tilde \e_i Y_i^3)  p_{1,\alpha} p_{2,\beta} \varepsilon_{\alpha\beta\mu\sigma} +\sum_\n p_{1,\nu} \Big(\hat\RR_{\mu,\nu,\sigma}({0}, 
{ 0})\nn\\
 +\sum_{a=1,2}
\sum_{\rho} p_{a,\rho} 
\frac{\partial \hat\RR_{\mu,\nu,\sigma}}{\partial p_{a,\rho}}({0}, {0}) \Big)=0
\ee
This implies that \be \hat\RR_{\mu,\nu,\sigma}({0}, 
{ 0})=0\ee and
\be
{\partial \hat\RR_{\mu,\nu,\sigma}\over \partial p_{2,\beta}} = - {1\over 6\pi^{2}}\varepsilon_{\nu\beta\mu\sigma} (\sum_i \tilde \e_i Y_i^3)\quad\quad
{\partial\hat\RR_{\mu,\nu,\sigma}\over \partial p_{1,\b}} (0, {0}) = {1\over 6\pi^{2}}\varepsilon_{\n\beta\mu\s} (\sum_i \tilde \e_i Y_i^3)
\ee
Finally using such values we get
\bea
&&\sum_\m (p_{1,\mu} + p_{2,\mu})\hat\Pi_{\mu,\n,\s}(p_1,p_2)=\sum_{\a,\b}\frac{(\sum_i \tilde \e_i Y_i^3)}{6\pi^{2}} p_{1,\alpha} p_{2,\beta} \varepsilon_{\alpha\beta\nu\sigma}\\
&&+\sum_{\mu,\b}  
(p_{1,\mu} + p_{2,\mu})
({\hat \RR_{\mu,\nu,\sigma}\over \partial p_{2,\beta}}(0,0) p_{2,\beta}
+{\hat \RR_{\mu,\nu,\sigma}\over \partial p_{1,\beta}} (0,0)
p_{1,\beta})\nn
\eea
and the second term in the r.h.s. is 
\be
-\frac1{6\pi^2} (p_{1,\mu} + p_{2,\mu}) \sum_{a=1,2}(-1)^a  p_{a,\beta}\varepsilon_{\nu\beta\mu\sigma}(\sum_i \tilde \e_i Y_i^3)\ee
 which is equal to
\be
\frac1{3\pi^2} 
p_{1,\mu} p_{2,\beta}\varepsilon_{\nu\beta\mu\sigma}(\sum_i \tilde \e_i Y_i^3)
\ee
which implies the result \pref{33aa}. \qed

Note that, in contrast to what happens in condensed matter problems 
\cite{BGM1}, the renormalizations depend on the chirality
and kind of particles and the anomaly itself is sum of different terms
which can cancel is a suitable condition is ensured.

\section{Appendix I. Truncated expectations}

For completeness we recall the proof of \pref{Br1}, referring for more details to App. B of \cite{Br} (see also
\cite{Mbook}, \cite{Ben}).
If 
$X=(1,2,..,n)$,  we call $V(X)=\sum_{i,j\in X} \bar V_{i,j}=\sum_{i\le j}V_{i,j}$ 
with $\bar V_{i,i}=V_{i,i}$ and $V_{i,j}=(\bar V_{i,j}+\bar V_{j,i})/2$ symmetric.
The starting point is the following formula
\be
e^{-V(X)}=\sum_{Y\subset X} K(Y) e^{-V(X/Y)}\label{iss1}
\ee
with $Y=X_1$, $|X_1|=1$ then $K(X_1)=e^{-V(X_1)}$ and for $r\ge 2$
\be
K(X_r)=\sum_T[\prod_{l\in T} V_l] 
[\sum_{r=2}^{n-1}  \sum_{X_1,..,X_{r-1} }\int_0^1 dt_1...\int_0^1 dt_{r-1}
{\prod_{k=1}^{r-1} t_k(l)\over t_{n(l)}} e^{-W_{X_r}(X_1,..,X_{r-1};t_1,..,t_{r-1})}] 
\label{issaa}
\ee
where $T$ is a tree connecting the points $(1,..,n)$, $l$ are bonds in $T$,
$X_1\subset X_2\subset...X_{r-1}$
are sets such that $|X_i|=i$ and each boundary $\partial X_i$ is crossed by at least a $l\in T$,
%\be
%K(X_r)=
%[\sum_{X_1,..,X_{r-1}}\sum_T [\prod_{l\in T} V_l] \int_0^1 dt_1...\int_0^1 dt_{r-1}
%\prod_{l\in T}{\prod_{k=1}^{r-1} t_k(l)\over t_{n(l)}} e^{-W_{X_r}(X_1,..,X_{r-1};t_1,..,t_{r-1})}] 
%\label{iss}
%\ee
%
%$n(l)$ is the max over $k$ such that $l$ crosses $\partial X_k$ and
%
\be
W_X(X_1,..,X_r;t_1,..,t_r)=\sum_{l} t_1(l)t_2(l)...t_r(l) V_l\label{ben11}
\ee
with $t_i(l)=t_i$ if $l$ crosses $\partial X_i$ and $t_i(l)=1$ otherwise, $n(l)$ is the max over $k$ such that $l$ crosses $\partial X_k$.
In order to prove \pref{iss1} we start noting that we can reverse the sum over $T$ and $X$
\be
\sum_T \sum_{X_1,..,X_{r-1}}=\sum_{X_1,..,X_{r-1}}\sum_T
\ee
where in the r.h.s. $T$ is a tree composed by $r-1$  lines $l$ such that all the
boundaries $\partial X_k$ are intersected at least by a line $l$.
We write by \pref{ben11}, if $X_1=\{1\}$
\be
W_X(X_1;t_1)=\sum_{\ell} t_1(l) V_l
\ee
with $ t_1(l)=t_1$ if $l$ crosses $\partial X_1$ and $=1$ otherwise; that is
\bea
&&W_X(X_1,t_1)=V_{1,1}+t_1 \sum_{k\ge 2} V_{1,k}
+\sum_{2\le k\le k'} V_{k,k'}=\nn\\
&&t_1 (V_{1,1}+\sum_{k\ge 2} V_{1,k}  +\sum_{2\le k\le k'} V_{k,k'})
+(1-t_1)(V_{1,1}+\sum_{2\le k\le k'} V_{k,k'})   =\nn\\
&&
t_1 V(X)+(1-t_1) (V(X_1)+V(X/X_1))\eea
We get $W_X(X_1,0)=V(X_1)+V(X/X_1)$ and
$
\partial_1 W(X_1,t_1)=\sum_{k\ge 2} V_{1,k}=\sum_{l_1} V_{l_1}$ so that
\be
e^{-V(X)}=\int_0^1 dt_1 \partial_1 e^{-W_X(X_1,t_1)}+e^{-W_X(X_1,0)}=\int_0^1 
dt_1 \sum_{k\ge 2} V_{1,k} e^{-W_X(X_1,t_1)}+ e^{-V(X_1)} e^{-V(X/X_1)}\label{tri}
\ee
Therefore $e^{-V(X)}$ is decomposed in the sum of two terms; in the first there is a bond  $(1,k)$ between
$X_1$ and the rest, in the second $X_1$ is decoupled.
If $n=2$ coincides with \pref{iss1}, $Y=X_1, X_2$ with $X_2=X$,
$e^{-V(X/X_2)}=1$.

If $X_2\not= X$ we further decompose the first term in the r.h.s of \pref{tri};
we write $X_2=\{1,k\}$ and
\bea
&&\int_0^1 dt_1 \sum_{k\ge 2} V_{1,k} 
e^{-W_X(X_1,t_1)}=\\
&&\int_0^1 dt_1 \sum_{k\ge 2} V_{1,k}
\int_0^1 dt_2 \partial_{t_2} e^{-W_X(X_1,X_2;t_1,t_2)}+\int_0^1 dt_1 \sum_{k\ge 2} V_{1,k}
e^{-W_X(X_1,X_2;t_1,0)}\nn
\eea
where
\be
W_X(X_1,X_2,t_1,t_2)=(1-t_2)[ W_{X_2}(X_1,t_1)+V(X/X_2)] +t_2 W_X(X_1,t_1)
\ee
If for instance $X_2=\{1,2\}$ then
\be
W_X(X_1,X_2,t_1,t_2)=
V_{1,1}+V_{2,2}+  t_1 t_2 \sum_{k\ge 3} V_{1,k}+
t_1 V_{1,2}
+t_2
 \sum_{k\ge 3} V_{2,k}
+\sum_{3\le k\le k'}  V_{k,k'}
\ee
%
%and 
%\bea
%&&W_X(X_1,X_2,t_1,1)=V_{1,1}+V_{2,2}+  t_1 \sum_{k\ge 3} V_{1,k}+t_1 V_{1,2}+
%\sum_{k\ge 3} V_{2,k}+\sum_{3\le k\le k'}  V_{k,k'}=W_X(X_1,t_1)\nn\\
%&&W_X(X_1,X_2,t_1,0)=V_{1,1}+V_{2,2}+t_1 V_{1,2}+\sum_{3\le k\le k'}  V_{k,k'}
%=W_{X_2}(X_1,t_1)+V(X/X_2)]
%\eea
%
Suppose that $X=\{1,2,3\}$ and $X_2=\{1,2\}$, then
\bea
&&\int_0^1 dt_1 V_{1,2} 
e^{-W_X(X_1,t_1)}=\\
&&\int_0^1 dt_1 V_{1,2}
\int_0^1 dt_2 (t_1 V_{1,3}+V_{2,3})
 e^{-W_X(X_1,X_2;t_1,t_2)}+[\int_0^1 dt_1 V_{1,2}
e^{-W_{X_2}(X_1;t_1)}]e^{-V(X/X_2)}\nn
\eea
The first term is $Y=X_3$
and the trees are $l_1=(1,2), l_2=(2,3)$ so that 
$t_1(l_1)=t_1$, $t_1(l_2)=1$, $t_2(l_2)=t_2$; 
and $l_1=(1,2)$, $l_2=(1,3)$ so that $t_1(l_1)=t_1$ and 
$t_1(l_2)=t_1$,$ t_2(l_2)=t_2$;  the second $Y=X_2$
and $K(X_2)e^{-V(X/X_2)}$.
If $X$ is larger than $X_3$, we further proceed
\be
e^{-W_X(X_1,X_2,t_1,t_2)}
=\int_0^1 dt_3 \partial_{t_3}   e^{-W_X(X_1,X_2,X_3t_1,t_2.t_3)  }+e^{-W_{X_3}(X_1,X_2,t_1,t_2)} e^{-V(X/X_3)}
\ee
and in particular if $X_3=(1,2,3)$ one writes
\bea
&&W(X_1,X_2,X_3,t_1,t_2,t_3)=t_1 t_2 V_{1,3}+t_2 V_{2,3}+\\
&& V_{1,1}+V_{2,2}+V_{3,3}+
t_1 t_2 t_3 \sum_{k\ge 4} V_{1,k}+t_2 t_3 \sum_{k\ge 4} V_{2,k}+
t_3 \sum_{k\ge 4} V_{3,k}+\sum_{k,k'\ge 4} V_{k,k'}\nn
\eea
If $X=X_4=\{1,2,3,4\}$ we get
\be
\int_0^1 dt_1 V_{1,2}
\int_0^1 dt_2 (t_1 V_{1,3}+V_{2,3}) \int_0^1 dt_3
 (t_1 t_2 V_{1,4}+t_2  V_{2,4}+V_{3,4})e^{-W_X(X_1,X_2,X_3t_1,t_2.t_3)  }
\ee
and the trees $T$ are; $l_1=(1,2), l_2=(1,3), l_3=(1,4)$ with $t_1(l_1)=t_1$, $t_1(l_2)=t_1, t_2(l_2)=t_2, 
 t_1(l_3)=t_1, t_2(l_3)=t_2, t_3(l_3)=t_3$; $l_1=(1,2), l_2=(1,3), l_3=(2,4)$,
with $t_1(l_2)=t_1$, $t_1(l_3)=1, t_2(l_3)=t_2, t_3(l_3)=t_3$; $l_1=(1,2), l_2=(2,3), l_3=(1,4)$ with
$t_1(l_2)=1,t_2(l_2)=t_2$
and so on. They are all possible trees compatible with $X_1=1, X_2=1,2,X_3=1,2,3$; then one has to sum over all the possible $X_i$.
Proceeding in this way one gets \pref{iss1}. 

Note finally that, see e.g. \cite{Br} or \cite{Mbook}
\be
\sum_{X_1,..,X_{r-1}}\int_0^1 dt_1...\int_0^1 dt_{r-1}
{\prod_{k=1}^{r-1} t_k(l)\over t_{n(l)}} =1
\ee
In order to prove \pref{for} we note that
by iterating \pref{iss1} we get
\be
e^{-V(X)}=\sum_\pi\prod_{Y\in \pi}
K(Y)\label{iss2}
\ee
where $\pi$ us the set of all possible partitions of $X$; on the other hand 
from \pref{dl1} we have $\EE_A(X)=e^{-V(X)}$
and
\be
\EE_A(X)=\sum_\pi\prod_{Y\in \pi}
\EE^T(Y)
\ee
hence $K=\EE^T$ from which \pref{for} follows.

Regarding the truncated fermionic expectations \pref{xx} we proceed in a similar way. 
We can write the simple expectations as
\be
\EE(\tilde\psi(P_1)...\tilde\psi(P_r))=\int \prod d\h_{i,j} e^{\sum_{j,j'} V_{jj'}  }
\ee
with $V_{jj'}=\sum_{i=1}^{|P_j|}\sum_{i'=1}^{|P_j'|} \h^+_{x_{ij} } g(x_{ij}, x_{i'j'})  \h^-_{x_{i'j'}}$
and $\h_{i,j}$ is a set of Grassmann variables. We can write therefore 
\bea
&&\EE(\tilde\psi(P_1)...\tilde\psi(P_r))=\\
&&\int \prod d\h_{i,j} \sum_T[\prod_{l\in T} V_l][\sum_{r=2}^{n-1}  \sum_{X_1,..,X_{r-1} }\int_0^1 dt_1...\int_0^1 dt_{r-1}
{\prod_{k=1}^{r-1} t_k(l)\over t_{n(l)}} e^{-W_{X_r}(X_1,..,X_{r-1};t_1,..,t_{r-1})}] \nn
\eea
with  $V_{j,j'}=\sum_{i}\sum_{i'} \h^+_{i,j} g(x_{ij},x_{i'j' }  \h^-_{i,j}$; for each $T$
we divide the $\h$ in the ones appearing in $T$ and $\h'$ is the rest; note that
\be
\int d\h_2 e^{  \sum_{l} t_1(l)t_2(l)...t_r(l) V_l}=\det G
\ee
where we have used that
\be
\sum_{l} t_1(l)t_2(l)...t_r(l) V_l=\sum_{l} t_{n'(l)}...t_{n(l)-1} (l) V_l
\ee
where $n(l)$ is the max $k$ such that $l$ intersects $\partial X_k$ and
$n'(l)$ is the min $k$ such that $l$ intersects $\partial X_k$; moreover 
$G$ is the matrix with elements $t_{n'(jj')}...t_{n(jj')-1} g(x_{i,j}, x_{i',j'} )$; 
for each sequence $X_k$ we can relabel the points so that $X_1=\{1\}$, $X_2=\{1,2\}$ and so on;
therefore given a line $j,j'$ then $n'(jj')=j, n(jj')=j'$ so that
$t_{j}...t_{j'-1}$; one can find a family of vectors $u_1=v_1, u_2=t_1 u_1+v_1 \sqrt{1-t_1^2},
u_3=t_2 u_2+v_2 \sqrt{1-t_2^2},...$$v_i$ orthonormal, such that 
$t_{j}...t_{j'-1}=u_j u_{j'} $.
\vskip.3cm
\noindent
{\it Acknowledgements} The work has been partly done as member of the Institute of Advanced Study in Princeton which we thank for support. We got support also from MUR, project MaQuMA,
PRIN201719VMAST01 and INDAM-GNFM.

\end{document}